\documentclass[twocolumn,preprintnumbers,superscriptaddress,altaffilletter,amsmath,amssymb]{revtex4}

\usepackage{graphicx}
\usepackage{dcolumn}
\usepackage{bm}
\usepackage{epsfig,amsfonts,amstext,afterpage,psfrag}
\usepackage{slashed}

\usepackage[latin1]{inputenc}

\def\bea{\begin{eqnarray}} 
\def\eea{\end{eqnarray}}
\def\be{\begin{equation}} 
\def\ee{\end{equation}}

\newcommand{\intfin}{\int_0^{z_0}}
\def\cA{{\cal A}}

\def\tr{\mathop{\rm tr}}

\newcommand{\beq}{\begin{equation}}
\newcommand{\eeq}{\end{equation}}

\newcommand{\tmfloatcontents}{}
\newlength{\tmfloatwidth}

\newcommand{\tmfloat}[5]{
	\renewcommand{\tmfloatcontents}{#4}
	\setlength{\tmfloatwidth}{\widthof{\tmfloatcontents}+1in}
	\ifthenelse{\equal{#2}{small}}
	{\ifthenelse{\lengthtest{\tmfloatwidth > \linewidth}}
		{\setlength{\tmfloatwidth}{\linewidth}}{}}
	{\setlength{\tmfloatwidth}{\linewidth}}  \begin{minipage}[#1]{\tmfloatwidth}
		\begin{center}
			\tmfloatcontents
			\captionof{#3}{#5}
		\end{center}
\end{minipage}}

\allowdisplaybreaks
\begin{document}
\preprint{SI-HEP-2019-19}

\title{Axial-vector and pseudoscalar mesons in the hadronic light-by-light contribution to the muon $(g-2)$} 
\author{Luigi Cappiello}
\affiliation{Dipartimento di Fisica \lq\lq Ettore Pancini'', Universit\'a di Napoli 'Federico II', Via Cintia, 80126 Napoli, Italy}
\affiliation{INFN-Sezione di Napoli, Complesso Universitario di Monte S. Angelo, Via Cintia Edificio 6, 80126 Napoli, Italy}

\author{Oscar Cat\`a}
\affiliation{Theoretische Physik 1, Universit\"at Siegen, Walter-Flex-Stra\ss e 3, D-57068 Siegen, Germany}

\author{Giancarlo D'Ambrosio}
\affiliation{INFN-Sezione di Napoli, Complesso Universitario di Monte S. Angelo, Via Cintia Edificio 6, 80126 Napoli, Italy}

\author{David Greynat}
\affiliation{No affiliation}

\author{Abhishek Iyer}
\affiliation{INFN-Sezione di Napoli, Complesso Universitario di Monte S. Angelo, Via Cintia Edificio 6, 80126 Napoli, Italy}
\affiliation{Universit\'e de Lyon, Universit{\' e} Claude Bernard Lyon 1, CNRS/IN2P3, UMR5822 IP2I,\\ F-69622, Villeurbanne, France}

\begin{abstract}
\begin{center} {\bf Abstract}\end{center}
Despite recent developments, there are a number of conceptual issues on the hadronic light-by-light (HLbL) contribution to the muon $(g-2)$ which remain unresolved. One of the most controversial ones is the precise way in which short-distance constraints get saturated by resonance exchange, particularly in the so-called Melnikov-Vainshtein (MV) limit. In this paper we address this and related issues from a novel perspective, employing a warped five-dimensional model as a tool to generate a consistent realization of QCD in the large-$N_c$ limit. This approach differs from previous ones in that we can work at the level of an effective action, which guarantees that unitarity is preserved and the chiral anomaly is consistently implemented at the hadronic level. We use the model to evaluate the inclusive contribution of Goldstone modes and axial-vector mesons to the HLbL. We find that both anomaly matching and the MV constraint cannot be fulfilled with a finite number of resonances (including the pion) and instead require an infinite number of axial-vector states. Our numbers for the HLbL point at a non-negligible role of axial-vector mesons, which is closely linked to a correct implementation of QCD short-distance constraints. 
\end{abstract}
\maketitle
\allowdisplaybreaks

\section{Introduction}\label{sec:1}

The anomalous magnetic moment of the muon is one of the most precise tests of the Standard Model dynamics. Besides the dominant electromagnetic contribution, one is also testing the weak and strong interactions. The present experimental value is given by $a_{\mu}^{\mathrm{exp}}=116 592 091(54)(33)\times10^{-11}$  \cite{pdg:2018}, where statistical errors are the largest source of uncertainty. The upcoming experiments at FNAL \cite{Venanzoni:2014ixa} and J-PARC \cite{Otani:2015lra} are expected to reduced the experimental error by a factor four down to $1.6\times 10^{-10}$, much smaller than the current theoretical uncertainty. 

If the E821 experimental number is confirmed, the discrepancy between the experimental value and the theoretical prediction, currently at about $3.5\sigma$, would rise up to a $7\sigma$ effect with the projected new precision. It is therefore essential to have good control over the theoretical estimate.

The theoretical prediction for $a_{\mu}$ is overwhelmingly dominated by electromagnetic~\cite{Aoyama:2012wk} and, to a much lesser extent, weak~\cite{Czarnecki:2002nt,Gnendiger:2013pva} effects (see also the reviews \cite{Jegerlehner:2009ry,Jegerlehner:2017gek}). Hadronic effects have a very modest contribution but are extremely difficult to evaluate. The present theoretical number, $a_{\mu}^{\mathrm{SM}}=116591823(1)(34)(26)\times 10^{-11}$ \cite{pdg:2018}, is dominated by the hadronic uncertainties (second and third error sources). The largest hadronic contribution comes from  the hadronic vacuum polarization, which can be rather cleanly connected to existing data on $e^+e^-$ scattering~\cite{Keshavarzi:2018mgv,Davier:2019can,Keshavarzi:2019abf}. In contrast, the (subleading) hadronic light-by-light contribution is more remote from experiment.  

The physics involved in the hadronic light-by-light (HLbL) contribution is sensitive to nonperturbative hadronic dynamics and cannot be calculated from first principles, except in some particular kinematical limits. One is therefore bound to use nonperturbative techniques. General arguments, based on chiral symmetry and the large-$N_c$ limit, can be used to assess the relevance of the different contributions \cite{deRafael:1993za}. There is general consensus that the neutral pion exchange provides the largest effect, and there is overall agreement on the contribution of the neutral Goldstone bosons. However, the status of the remaining contributions is not as satisfactory, in particular that of axial-vector mesons. Depending on the method used, their estimated contribution can differ by one order of magnitude, from being negligible to accounting for roughly 15\% of the value of the HLbL. The latter value is comparable with the projected experimental precision at FNAL and J-PARC, so a better understanding of the axial-vector contribution is definitely needed.

Progress on the light-by-light front is nowadays pursued along three main avenues: hadronic models, dispersion relation approaches and lattice simulations. 

Hadronic models provide, by far, the largest pool of HLbL determinations. In some cases the models are rather broad in scope \cite{Bijnens:1995cc,Bijnens:1995xf,Hayakawa:1995ps,Hayakawa:1996ki}, while in some other cases \cite{Hayakawa:1997rq,Knecht:2001qf,Melnikov:2003xd} the focus is on specific contributions. The main strategy behind these approaches is to come up with hadronic form factors able to successfully interpolate between low energies, where experimental data is available, and high energies, where the OPE of QCD is valid. The different models can then be understood as different ways to build interpolating functions between these limiting cases. In principle, the more constraints that a model satisfies, the more reliable their predictions should be. This has motivated a lot of work to increase the number of known short-distance constraints and test their impact on $a_{\mu}$ \cite{Melnikov:2003xd,Nyffeler:2009tw,Bijnens:2019ghy}. 

Dispersion relation techniques have been applied more recently (see, e.g.,~\cite{Colangelo:2014dfa,Colangelo:2014pva,Colangelo:2015ama,Pauk:2014rfa}). Their main focus is to bring the HLbL determination as close as possible to the available experimental data, thus affording much better control of the uncertainties with respect to hadronic models. Progress is underway but there are still a number of open issues. In particular, how dispersive techniques should implement the short-distance constraints from perturbative QCD is only starting to be studied (see, e.g.,~\cite{Colangelo:2019lpu,Colangelo:2019uex,Melnikov:2019xkq,Knecht:2020xyr}). 

In turn, lattice simulations are rapidly becoming competitive \cite{Blum:2016lnc,Blum:2017cer,Asmussen:2018oip,Blum:2019ugy} and should eventually give us the most precise determination of the muon HLbL. 

However, the effort to bring the HLbL under better theoretical control also requires the resolution of a number of conceptual issues that are still open. The core of the problem is to understand how short distances are saturated by the different hadronic states. In the HLbL, understanding how these duality relations work turns out to be a highly nontrivial task. A notorious example is the short-distance constraint discussed in \cite{Melnikov:2003xd}, which was claimed to increase substantially the value for the HLbL contribution through a combined increase of the Goldstone and axial-vector contributions. Attempts to incorporate the constraint into form factor models have led to a number of proposals, e.g.,~\cite{Melnikov:2003xd,Jegerlehner:2009ry}. However, without a better understanding of how this constraint happens to be fulfilled, the state of affairs with the HLbL cannot be considered satisfactory.

In order to address the previous point, it is clear that one has to go beyond form factor parametrizations and be able to compute in terms of hadronic states at the level of correlators. This can be done if one borrows techniques from QFT in extra dimensions. It is well-known that, starting from a five-dimensional theory, the compactification to four dimensions gives rise to an infinite number of modes, which can be interpreted as mesons \cite{Sakai:2004cn,Erlich:2005qh,Hirn:2005nr,DaRold:2005zs}. These constructions can also be tailored to break chiral symmetry spontaneously, such that in the infrared limit one recovers chiral perturbation theory. If, additionally, the five-dimensional theory lives in an anti de-Sitter (AdS) gravitational background, the breaking of the associated conformal symmetry mimics the almost conformal behavior of QCD at large momenta. As a result, these theories approximate remarkably well both the long- and short-distance behavior of QCD correlators. Finally, if the five-dimensional model is endowed with a Chern-Simons term, one obtains a four-dimensional theory with the chiral anomaly consistently implemented at all energy scales (see, e.g.,~\cite{Hill:2006wu,Panico:2007qd}). Following the AdS/CFT prescription \cite{Maldacena:1997re,Gubser:1998bc,Witten:1998qj}, there is a well-defined procedure to compactify the fifth dimension and express the resulting four-dimensional effective action in terms of external sources, out of which correlators can be computed with functional differentiation. The five-dimensional model is therefore used here simply as a technical device to end up with a consistent four-dimensional theory of hadrons.

Similar five-dimensional settings have been used to evaluate the Goldstone~\cite{Hong:2009zw,Cappiello:2010uy,Leutgeb:2019zpq} and axial-vector contributions~\cite{Leutgeb:2019gbz} to the HLbL. In this paper we employ the simplest model implementing all of the features mentioned above to evaluate the joint Goldstone and axial-vector contributions to the HLbL correlator. As opposed to other approaches, our determination considers all the states coupled to axial currents in an inclusive way and unitarity is thus automatically built in. Considering Goldstone and axial-vector states simultaneously is also a requisite to fulfill anomaly matching, as we will show below.

Our analysis clarifies a number of points. First, it shows that the MV condition can only be fulfilled by a collective effect of the axial-vector resonances and not by a finite number of form factors. Similar results were found in~\cite{Leutgeb:2019gbz} and we thus confirm their conclusions. This explains, in particular, why attempts to fulfill the condition with single-particle form factors were problematic, no matter their degree of sophistication. Second, this collective axial-vector effect is intimately connected with having anomaly matching at all energies. With a finite number of axial-vector mesons, anomaly matching simply fails. This shows that axial-vector mesons have a more prominent role than previously assumed. 

Our number for the {\it{joint}} pseudoscalar and axial-vector contribution to the HLbL is $a_{\mu}=125(15)\cdot 10^{-11}$, largely compatible with the numbers of other hadronic models that incorporate the MV constraint. However, we argue that part of the axial-vector contribution was previously misidentified as coming from pseudoscalars. As a result, we find that the relative weight of the axial-vector contribution onto the HLbL is larger than previously claimed.

The structure of this paper is as follows. In Sec.~\ref{sec:2} we introduce our toy model and bring it to the form of a four-dimensional effective action. We highlight how the pion and the (infinite towers of) vector and axial-vector resonances arise. In Sec.~\ref{sec:3} we derive the expression for the HLbL electromagnetic tensor as predicted by the model in a closed form. The HLbL tensor is then naturally split in two pieces, which collect the longitudinal and transverse polarizations of the resonances exchanged. The structure of the longitudinal piece is explored in detail in Sec.~\ref{sec:4}, and in Sec.~\ref{sec:5} we show its relation with the triangle anomaly by analyzing the $VVA$ correlator. The MV short-distance constraint is discussed in Sec.~\ref{sec:6}. We show explicitly how the model fulfills it and make the connection with anomaly matching. In Sec.~\ref{sec:7} we give our numbers for the Goldstone and axial-vector contributions to the muon anomalous magnetic moment and compare them with previous estimates. Concluding remarks are given in Sec.~\ref{sec:8}, while technical aspects are collected in three Appendices.  

\section{The model}\label{sec:2}

In order to have a consistent realization of hadronic physics in the large-$N_c$ limit for vectors, axial-vectors and Goldstone bosons, we will adopt an extension of the five-dimensional model introduced in \cite{Hirn:2005nr}, which is a particular application of the AdS/CFT correspondence \cite{Maldacena:1997re} to hadronic physics. The spirit of using this model is to have a minimal setup able to capture the relevant features of QCD for the HLbL, namely conformality at very high energies, chiral symmetry breaking at low energies, and the chiral anomaly. The model can also be extended to incorporate scalars and (non-Goldstone) pseudoscalars, but this further step will not be considered in the present paper.
   
The model is a five-dimensional $U(3)_L\times U(3)_R$ Yang-Mills--Chern-Simons theory, 
\begin{align}\label{SYM5}
S&=-\lambda\int d^5x\sqrt{g}\,{\rm tr}\!\left[F_{(L)}^{MN}F_{(L)MN}+F_{(R)}^{MN}F_{(R)MN}\right]\nonumber\\
&+c \int d^5x \,\tr\big[\omega_5(L)-\omega_5(R) \big]\,,
\end{align}
where $L_M=L_M^at^a$ is a $U(3)_L$ gauge field, $F_{(L)\,MN} =\partial_{M}L_{N}-\partial_{N}L_{M}-i[L_{M},L_{N}]$ and
\begin{equation}\label{omega5}
\omega_5(L)= \tr\left[ L F_{(L)}^2+\frac{i}{2}
{L}^3F_{(L)}-\frac{1}{10} L^5 \right]\,,
\end{equation}
where the wedge product of forms is implicitly understood. Similar considerations apply to the right-handed sector. $t^a$ are the eight Gell-Mann matrices extended with $t^0=\textbf{1}_3/\sqrt{6}$, normalized such that ${\mathrm{tr}}\,(t^a \,t^b)=\tfrac{1}{2}\delta_{ab}$. 

A point in the five-dimensional space has coordinates $(x,z)$. The background metric will be chosen to be exactly anti de-Sitter, so that
\begin{align}
g_{MN}dx^Mdx^N&=\frac{1}{z^2}\left(\eta_{\mu \nu}dx^{\mu}dx^{\nu} -
dz^2\right)\,,
\end{align} 
where $\mu,\nu=(0,1,2,3)$, $M,N=(0,1,2,3,z)$ and $\eta_{\mu\nu}$ has a mostly negative signature. The fifth dimension is assumed to be compact, i.e., four-dimensional boundary branes exist at $(x,0)$ and $(x,z_0)$, the so-called UV and IR boundary branes, respectively. 

In order to make contact with the hadronic states, it is convenient to trade the left- and right-handed fields, $L_{\mu}(x,z)$ and $R_{\mu}(x,z)$, for vector and axial-vector ones through the usual relations $L_{\mu}=V_{\mu}-A_{\mu}$ and $R_{\mu}=V_{\mu}+A_{\mu}$. These (massless) fields admit Kaluza-Klein decompositions
\begin{align}
V_{\mu}(x,z)&=\sum_{n}V_{\mu}^{(n)}(x)\varphi_n^V(z)\,,\nonumber\\
A_{\mu}(x,z)&=\sum_{n}A_{\mu}^{(n)}(x)\varphi_n^A(z)\,,
\end{align}
and generate two infinite towers of four-dimensional modes, which become massive by absorbing the scalar modes $V_5^{(n)}$ and $A_5^{(n)}$ through higgsing. 

The resonance poles are determined from the solutions for $\varphi_n^{V,A}(z)$. Working in four-dimensional momentum space, they are normalizable only for discrete values of the four-momentum $q$, namely at
\begin{align}\label{spectrum}
m_{Vn}=\frac{\gamma_{0,n}}{z_0}\,;\quad m_{An}=\frac{\gamma_{1,n}}{z_0}\,,
\end{align}
where $\gamma_{k,n}$ is the $n^{th}$ root of the Bessel function $J_k(x)$. The previous equation shows that the size of the fifth dimension is an infrared quantity that sets the confinement scale. 

Spontaneous chiral symmetry breaking is implemented in this model through boundary conditions on the IR brane, where low-energy physics takes place. The choice 
\begin{align}
L_{\mu}(x,z_0)-R_{\mu}(x,z_0)=0\,,\nonumber\\
F_L^{z\mu}(x,z_0)+F_R^{z\mu}(x,z_0)=0\,,
\end{align}
ensures that on the infrared brane only the vectorial subgroup $U(3)_V$ is preserved. The pattern of breaking is therefore the one expected from large-$N_c$ QCD, namely $U(3)_L\times U(3)_R\to U(3)_V$, and a nonet of Goldstone bosons is generated. 

The infrared boundary conditions make sure that all the zero modes cancel except $A_5^{(0)}$, which encodes the Goldstone degrees of freedom. In order to have a more conventional representation of the pion multiplet, it is convenient to trade the $A_5^{(0)}$ field for a Wilson line. One defines
\begin{align}
 \xi_{L}(x,z)=P  \exp \left\{-i \int_z^{z_0} dz'\,
L_5(x,z') \right\}\,,\label{Wilson}
 \end{align}
and $\xi_{R}(x,z)$ in a similar way, such that the IR boundary conditions are respected, and redefines the fields as 
\begin{align}\label{fieldred}
L^{\xi}_{M}(x,z)=\xi_L^{\dagger}(x,z)\left[L_{M}(x,z) + i\partial_{M}\right]\xi_L(x,z)\,,\nonumber\\
R^{\xi}_{M}(x,z)=\xi_R^{\dagger}(x,z)\left[R_{M}(x,z) + i\partial_{M}\right]\xi_R(x,z)\,.
\end{align}
These chirally dressed combinations make sure that the physical degrees of freedom are $L^{\xi}_{\mu}(x,z)$ and $R^{\xi}_{\mu}(x,z)$, while their fifth components identically vanish. 

The Goldstone degrees of freedom are captured by the combination of Wilson lines $(\xi_{\cA}(x, 0) \equiv \xi_{\cA}(x))$: 
\beq
U(x)\equiv \xi_{L}(x)\xi_{R}^\dagger (x)=\exp\left[\frac{2i\pi^a(x)t^a}{f_\pi}\right]\,,\label{Uchiral}
\eeq
which transforms as $U(x)\to g_L(x)U(x)g_R^{\dagger}(x)$. The field redefinitions in (\ref{fieldred}) are thus the way to move from a linear to a nonlinear representation of chiral symmetry breaking. With $SU(3)_L\times SU(3)_R$, one can always choose $\xi_L(x)=\xi_R^{\dagger}(x)\equiv u(x)$, such that $U(x)=u^2(x)$ \cite{Callan:1969sn}. As a result, the expressions for the chirally dressed UV sources are
\begin{align}\label{dressed}
l^{\xi}_{\mu}\equiv L^{\xi}_{\mu}(x,0)=u^{\dagger}(x)\left[l_{\mu}(x) + i\partial_{\mu}\right]u(x)\,,\nonumber\\
r^{\xi}_{\mu}\equiv R^{\xi}_{\mu}(x,0)=u(x)\left[r_{\mu}(x) + i\partial_{\mu}\right]u^{\dagger}(x)\,.
\end{align}
The Yang-Mills action in eq.~(\ref{SYM5}) is invariant under this field redefinition. In contrast, the Chern-Simons form gets shifted to
\begin{align}
\omega_5(L^{\xi})=\omega_5(L)+\omega_5(\Sigma_L)+d\alpha_4(L,\Sigma_L)\,,
\end{align}
where $\Sigma_L=d\xi_L\xi_L^{\dagger}$, and similarly for the right-handed fields. The second term can be shown to reproduce the ungauged Wess-Zumino-Witten Lagrangian, while the function
\begin{align}\label{BC}
\alpha_4(L,\Sigma_L)=\frac{1}{2}{\mathrm{tr}}&\left[ \Sigma_L (LF_{(L)}+F_{(L)}L)\right.\nonumber\\
&\left.+i\Sigma_L L^3-\frac{1}{2}\Sigma_L L\Sigma_L L-i\Sigma_L^3L\right]
\end{align}
is a pure boundary term in five dimensions.

The connection with the associated effective four-dimensional theory is done with the AdS/CFT correspondence prescription~\cite{Gubser:1998bc, Witten:1998qj}, according to which the value of the five-dimensional fields on the UV brane are the sources of the four-dimensional operators. It is therefore convenient to split the fields as
\begin{align}\label{decomposition}
A_{\mu}(x,z)&=a(x,z){\hat{a}}^{\perp}_{\mu}(x)+{\bar{a}}(x,z){\hat{a}}^{\parallel}_{\mu}(x)+\frac{\alpha(z)}{f_{\pi}}\partial_{\mu}\pi(x)\,,\nonumber\\
V_{\mu}(x,z)&=v(x,z){\hat{v}}^{\perp}_{\mu}(x)+{\bar{v}}(x,z){\hat{v}}^{\parallel}_{\mu}(x)\,,
\end{align}
where ${\hat{v}}(x)$ and ${\hat{a}}(x)$ are identified with the classical sources associated to the chiral currents
\beq j^a_{\mu}= \overline{q} \gamma_{\mu}t^a
q,\quad j^{5a}_{\mu}= \overline{q}\gamma_{\mu}\gamma^5t^a q\,.
\label{chiralcurrents}\eeq
The functions $a(x,z)$, ${\bar{a}}(x,z)$, $ \alpha(z)$, $v(x,z)$ and ${\bar{v}}(x,z)$ can be found by solving the (linearized) five-dimensional equations of motion subject to the appropriate boundary conditions (see Sec.~\ref{sec:3}).

In order to obtain the four-dimensional effective action, the solutions in eq.~(\ref{decomposition}) are substituted into the five-dimensional action and the dependence on the fifth dimension is integrated out. The end result is a four-dimensional generating functional, out of which the correlators of the theory can be computed. 

The model presented here contains an infinite tower of vector and axial-vector resonances together with the pion multiplet and is therefore a good toy model to evaluate the interplay between Goldstone modes and axial vectors in the HLbL. 

All observables are expressed in terms of the three parameters $\lambda$, $c$ and $z_0$. The former is normally fixed by matching the coefficient of the parton logarithm in the axial-vector two-point function (see, e.g.,~\cite{Erlich:2005qh} and Appendix~\ref{app:2}). This gives
\beq\label{g5Nc}
\lambda=\frac{N_c}{48\pi^2}\,.
\eeq
By requiring the right normalization of the chiral anomaly, one finds
\beq\label{anomaly}
c=\frac{N_c}{24\pi^2}\,.
\eeq
The parameter $z_0$ is a characteristic infrared scale. Actually, the simplicity of the model means that all infrared quantities depend on $z_0$. For instance, the pion decay constant is given by
\beq
f_\pi^2=\frac{8\lambda}{z_0^2}=\frac{N_c}{6\pi^2z_0^2}\label{fpi}\,,
\eeq
and the resonance masses are given in eq.~(\ref{spectrum}). Which parameter is chosen to fix $z_0$ depends on the application at hand. We will come back to this issue in Sec.~\ref{sec:7}. For the time being, we simply observe that the model predicts extremely well the splitting between the lowest-lying vector and axial-vector states, i.e.,
\begin{align}
\frac{m_{\rho}}{m_{a_1}}=\frac{\gamma_{0,1}}{\gamma_{1,1}}\sim 0.63\,,\label{ratiom}
\end{align}
but cannot account for satisfactory values for $f_{\pi}$ and $m_{\rho}$ simultaneously. These shortcomings can be circumvented but at the price of sophisticating the model in a rather {\it{ad hoc}} way. In this paper we will stick to the minimal model. As we will argue in the next sections, since not all the hadronic information is equally relevant for the HLbL, the minimal model is already able to provide interesting quantitative estimates.

It is also relevant to mention at this point that the success of eq.~(\ref{ratiom}) does not extend to heavier vector and axial-vector mesons. There is also the issue of the mass splittings inside multiplets, something that the model is unable to capture. One could therefore cast some doubts on the reliability of the model to yield predictions for the HLbL. The key point is that the HLbL is an inclusive observable for {\emph{spacelike}} momenta. In these cases, practice has shown that correlators with very different spectra lead to very stable predictions, as long as short and long distances are correctly matched (see, e.g., the discussion in~\cite{deRafael:2002tj}). As a result, we expect our prediction for the HLbL contributions to be rather robust, despite the fact that the individual resonance contributions might be hard to match to the actual QCD hadronic spectrum.

Another limitation of the model is that the Goldstone modes are strictly massless, i.e., there is no explicit chiral symmetry breaking. In order to be realistic, we should depart from the chiral limit and give the pion multiplet a mass. We will account for explicit chiral symmetry breaking simply by adding a mass term in the Goldstone propagators, while the form factors will be computed in the chiral limit, which is known to be a very good approximation~\cite{Guevara:2018rhj}. This modification can be understood as introducing a deformation operator on the UV boundary and therefore does not jeopardize the consistency of the model. For more details see, for instance,~\cite{Brunner:2015oga,Bartolini:2016dbk} and references therein.  
 
\section{The electromagnetic four-point function}\label{sec:3}

\begin{figure}[t]
\begin{center}
\includegraphics[width=0.47\textwidth]{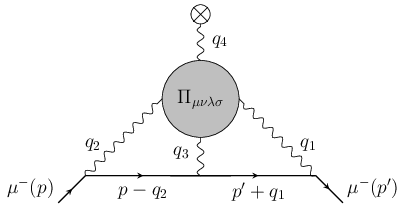}
\end{center}
\caption{The HLbL diagram. The blob represents the HLbL tensor. In our conventions, photon momenta are pointing inwards.} \label{fig:1}
\end{figure}

The fundamental object for the HLbL is the electromagnetic four-point correlator of fig.~\ref{fig:1}, defined as
\begin{align}
&\Pi^{\mu\nu\lambda\rho}(q_1,q_2,q_3) = -i \int d^4x\ d^4y\ d^4z \ e^{-i(q_1 \cdot x + q_2 \cdot y + q_3 \cdot z)}\nonumber\\
&\qquad\times \langle 0 | T \{ j_\mathrm{em}^\mu(x) j_\mathrm{em}^\nu(y) j_\mathrm{em}^\lambda(z) j_\mathrm{em}^\rho(0) \} | 0 \rangle\,, \label{HLbLTensorDefinition}
\end{align}
where $j_\mathrm{em}^\mu(x)={\bar{q}}\gamma^{\mu}{\hat{Q}}q$, with ${\hat{Q}}=\tfrac{1}{3}{\mathrm{diag}}(2,-1,-1)$ being the electromagnetic charge matrix. Our conventions for momenta are such that $q_1+q_2+q_3+q_4=0$.

This correlator satisfies the Ward identities:
\begin{align}
\left\{q_1^{\mu},q_2^{\nu},q_3^{\lambda},q_4^{\rho}\right\}\times\Pi_{\mu\nu\lambda\sigma}(q_1,q_2,q_3)=0\,,
\end{align}
which reduce the number of independent kinematic invariants down to 43 gauge-invariant tensor structures \cite{Colangelo:2015ama}.

In our model, all quartic terms in vector fields are antisymmetric in flavor indices and therefore cancel due to Bose symmetry. The leading contributions to eq.~(\ref{HLbLTensorDefinition}) are driven by the cubic interactions in the Chern-Simons term, corresponding to pion and axial-vector exchanges, which are the leading effects in the $1/N_c$ expansion. 


\begin{figure}[t]
\begin{center}
\includegraphics[width=0.48\textwidth]{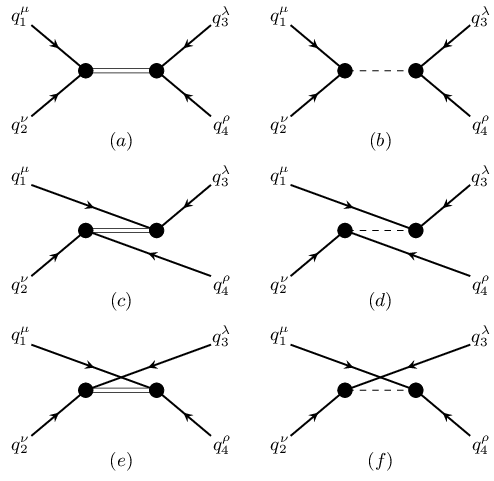}
\end{center}
\caption{Diagrams contributing to the HLbL tensor at tree level in our model. The solid lines represent the vector bulk-to-boundary propagators, depending on the external momenta. The double lines in $(a)$, $(c)$ and $(e)$ denote the axial-vector Green functions, while the dashed lines in $(b)$, $(d)$ and $(f)$ correspond to the pion propagator. The black dots represent the trilinear anomalous vertices, derived from the Chern-Simons part of the action. To any of these vertices an integration over the fifth dimension is understood.}\label{fig:2}
\end{figure}

The corresponding diagrams are listed in fig.~\ref{fig:2}. The vertices can be extracted from the effective action, which is obtained by solving the equations of motion for vector and axial-vector fields and plugging the solutions back into the five-dimensional action. As usual, exact solutions cannot be found and one has to resort to perturbation theory, with the quadratic part of the Yang-Mills piece as the leading effect and the Chern-Simons term as a perturbation.

The Yang-Mills piece of the action can be written in components as
\begin{align}\label{YM}
S_{\rm YM}&[V,A]=-2\lambda\int d^4x \int_0^{z_0} \frac{dz}{z}\nonumber\\
&\times{\mathrm{tr}}\left[(F^{V}_{\mu\nu})^2-2(F_{\mu z}^V)^2+(F^{A}_{\mu\nu})^2-2(F_{\mu z}^A)^2\right]\,.
\end{align}
From the Chern-Simons term we need to keep interactions linear in the axial-vector field. They come from the first terms in eqs.~(\ref{omega5}) and (\ref{BC}). In components, one finds 
\begin{align}\label{effaction}
&S_{{\mathrm{CS}}}^{(3)}[V,A]=2c\varepsilon^{\mu\nu\lambda\rho}{\hat{d}}^{abc}\left[-\frac{1}{2f_{\pi}}\int d^4x \pi^a\partial_{\mu}V_{\nu}^b\partial_{\lambda}V_{\rho}^c\right.\nonumber\\
&\left.\!\!\!+\int d^5x \left(A_{\mu}^a\partial_{\nu}V_{\lambda}^b\partial_z V_{\rho}^c-\partial_z A_{\mu}^a\partial_{\nu}V_{\lambda}^bV_{\rho}^c-\partial_{\nu}A_{\mu}^aV_{\lambda}^b\partial_z V_{\rho}^c\right)\right]\,,
\end{align}
where ${\hat{d}}^{abc}={\rm{tr}}\big[t^a\{t^b,t^c\}\big]$ and the first piece comes from the boundary term in eq.~(\ref{BC}). Plugging the expressions for the vector and axial-vector fields given in eq.~(\ref{decomposition}), the previous expression can be split into a transverse, longitudinal and the Goldstone contribution. In the following, we will concentrate on the contributions of the neutral pion and the $a_1(1260)$ tower, and we will select accordingly the flavor structure ${\hat{d}}^{3\gamma\gamma}=\tfrac{1}{3}$. For the numerical estimates in Sec.~\ref{sec:7}, we will also consider the contributions of the $\eta$, $\eta'$ and the axial-vector isosinglet towers. 

In order to build the diagrams of fig.~\ref{fig:2} from the action one needs the first-order solutions for vector fields and up to second-order solutions for the axial-vector ones. The first-order solutions are the so-called bulk-to-boundary propagators. Working in four-dimensional (Euclidean) momentum space, they read
\begin{align}
v(z,Q)&=Qz\left[K_1(Qz)+\frac{K_0(Qz_0)}{I_0(Qz_0)} I_1(Qz)\right]\,,\nonumber\\
a(z,Q)&=Qz\left[K_1(Qz)-\frac{K_1(Qz_0)}{I_1(Qz_0)} I_1(Qz)\right]\,,
\end{align}
where $K_j(x)$ and $I_j(x)$ are modified Bessel functions. The axial-vector zero-mode $\alpha(z)$ instead simplifies to 
\beq\label{alphaFunct}
\alpha(z)=1-\frac{z^2}{z_0^2}\,.
\eeq
The second-order solution for the axial-vector field can be expressed in terms of the first-order ones as 
\begin{align}
A_{\mu}^{(1)}(x,z)=\frac{c}{2\lambda}\epsilon^{\alpha\nu\lambda\rho}\int_0^{z_0} d\xi\, G^A_{\alpha\mu}(z,\xi;x)\partial_{\xi}V_{\nu}\partial_{\lambda}V_{\rho}\,,
\end{align}
where $G_A^{\mu\nu}(z,\xi;x)$ is the axial-vector Green function (see Appendix~\ref{app:1}). The expression for the transverse and longitudinal components can be easily found by projecting out the corresponding components of the Green function, defined as
\begin{align}\label{decomp1}
G_{\mu\nu}^A(z,z^{\prime};q)&= P^{\perp}_{\mu\nu}G_{\perp}^A(z,z^{\prime};q)+P^{\parallel}_{\mu\nu}G_{\parallel}^A(z,z^{\prime};q)\,,
\end{align} 
with
\begin{align}
P^{\perp}_{\mu\nu}=\eta_{\mu\nu}-\frac{q_{\mu}q_{\nu}}{q^2};\quad P^{\parallel}_{\mu\nu}=\frac{q_{\mu}q_{\nu}}{q^2}\,.
\end{align}

Plugging the previous solutions back into eq.~(\ref{SYM5}), the relevant terms of the effective action for the electromagnetic tensor are
\begin{widetext}
\begin{align}\label{VVVV}
S_{\rm{eff}}&\supset \int d^4x\left\{\frac{c^2}{\lambda}\varepsilon^{\mu\nu\lambda\rho}\varepsilon^{\mu'\nu'\lambda'\rho'}\intfin \intfin dz\, dz' \bigg[\partial_{\nu}V_{\lambda}(x,z)\partial_z V_{\rho}(x,z)\bigg] G^A_{\mu\mu'}(z,z^{\prime};x) \bigg[\partial_{\nu'}V_{\lambda'}(x,z')\partial_{z'} V_{\rho'}(x,z')\bigg]\right.\nonumber\\
&\left.+\frac{1}{2}\partial_{\mu}\pi(x)\partial^{\mu}\pi(x)+\frac{c}{f_{\pi}}\varepsilon_{\mu\nu\lambda\rho}\pi(x)\int_0^{z_0}dz\,\alpha'(z)\partial_{\mu}V_{\nu}(x,z)\partial_{\lambda}V_{\rho}(x,z)\right\}\,.
\end{align}
\end{widetext}

The first line takes into account the contribution of axial vectors while the second line contains the form factor of the pion coupled to two photons, which is defined as 
\begin{align}\label{pigamma2}
\Gamma_{\mu\nu}(q_1,q_2)&=i\int d^4 x \ e^{iq_1\cdot x } \langle 0|T\left\{j_{\mu }^{\mathrm{em}
}(x)\,j_{\nu}^{\mathrm{em}}(0)\right\}| \pi(p) \rangle\nonumber\\
&=\varepsilon_{\mu\nu\alpha\beta}q_{1}^{\alpha} q_{2}^{\beta} \,
F_{\pi\gamma\gamma} \left(q_1^2,q_2^2 \right )\,.
\end{align}
The pion propagator follows directly from the term
\begin{align}\label{Seff2}
-2\lambda A_{\mu}(x,z)\frac{1}{z}\partial_z A^{\mu}(x,z)\bigg|_{z=0}\,,
\end{align}
which is the effective action for axial-vector fields coming from eq.~(\ref{YM}), once the expression in eq.~(\ref{decomposition}) is used. Notice that the boundary term in (\ref{effaction}) affecting the pion is no longer present. Its cancellation, leaving the expression above for the pion form factor, is a consistency check that our model has the anomaly correctly implemented. The presence of a boundary term would be in conflict, e.g., with the asymptotic behavior of the form factor at large photon virtualities.
 
By matching (\ref{pigamma2}) to the holographic expression above, one finds 
\begin{align}\label{pigammagammaff}
F_{\pi\gamma\gamma}(Q_1^2, Q_2^2)&=\frac{2c}{f_{\pi}}\int_0^{z_0} dz\;\alpha'(z)\:v(z,Q_1)v(z,Q_2)\,.
\end{align}
The expression above depends on the three parameters of the model, which can be traded for $c$, $z_0$ and $f_{\pi}$. However, as opposed to other hadronic models, in our approach the functional form of $F_{\pi\gamma\gamma}$ is completely fixed, so eq.~(\ref{pigammagammaff}) is actually a consistency check of the model.

In the zero-momentum limit, $F_{\pi\gamma\gamma}$ is determined by the chiral anomaly. Using that $v(z,0)=1$, the integral above is given by the boundary values for $\alpha(z)$. From eq.~(\ref{anomaly}) one then obtains the well-known result 
\begin{align}
F_{\pi\gamma\gamma}(0,0)=-\frac{N_c}{12\pi^2 f_\pi}\,,
\end{align}
which confirms that the model has the chiral anomaly correctly implemented.

At very high energies, for large and equal photon momenta, one can expand eq.~(\ref{pigammagammaff}) to find 
\begin{align}\label{OPEpion}
\lim_{Q^2\to \infty}F_{\pi\gamma\gamma}(Q^2,Q^2)=-\frac{2f_{\pi}}{3Q^2}+{\cal{O}}\left(e^{-Qz_0}\right)\,,
\end{align}
which matches the OPE prediction. When one photon is on shell and the other far off shell, one expects the Brodsky-Lepage $Q^{-2}$ scaling. We indeed find 
\begin{align}\label{OPEBL}
\lim_{Q^2\to \infty}F_{\pi\gamma\gamma}(0,Q^2)=-\frac{2f_{\pi}}{Q^2}+{\cal{O}}\left(e^{-Qz_0}\right)\,.
\end{align}
Eq.~(\ref{pigammagammaff}) therefore has the correct high- and low-energy behavior. Notice, however, that only the leading term in the OPE is correctly reproduced by the model, with all subleading pieces identically vanishing. This is a consequence of the conformal symmetry of the AdS metric.  

The electromagnetic four-point function defined in eq.~(\ref{HLbLTensorDefinition}) can now be obtained by taking the variation of eq.~(\ref{VVVV}) with respect to the boundary values of the (transverse) vector fields. Using eq.~(\ref{pigammagammaff}), one obtains  
\begin{widetext}
\begin{align}\label{VVVVM}
\Pi_{\mu\nu\lambda\rho}(q_1,q_2,q_3,q_4)&=\varepsilon_{\mu\nu\alpha\beta}\varepsilon_{\lambda\rho\alpha'\beta'}\left[\frac{2c^2}{\lambda}\int dz \int dz' T^{\beta}_{12}(z)G_A^{\alpha\alpha'}(z,z^{\prime};s)T^{\beta'}_{34}(z^{\prime})+F_{\pi\gamma\gamma}(q_1,q_2)\frac{q_1^{\alpha}q_2^{\beta}q_3^{\alpha'}q_4^{\beta'}}{s-m_\pi^2}F_{\pi\gamma\gamma}(q_3,q_4)\right]\nonumber\\
&+\varepsilon_{\mu\nu\alpha\alpha'}\varepsilon_{\lambda\rho\beta\beta'}\left[\frac{2c^2}{\lambda}\int dz \int dz' T^{\beta}_{13}(z)G_A^{\alpha\alpha'}(z,z^{\prime};t)T^{\beta'}_{24}(z^{\prime})+F_{\pi\gamma\gamma}(q_1,q_3)\frac{q_1^{\alpha}q_2^{\beta}q_3^{\alpha'}q_4^{\beta'}}{t-m_\pi^2}F_{\pi\gamma\gamma}(q_2,q_4)\right]\nonumber\\
&+\varepsilon_{\mu\nu\alpha\beta'}\varepsilon_{\lambda\rho\beta\alpha'}\left[\frac{2c^2}{\lambda}\int dz \int dz' T^{\beta}_{14}(z)G_A^{\alpha\alpha'}(z,z^{\prime};u)T^{\beta'}_{23}(z^{\prime})+F_{\pi\gamma\gamma}(q_1,q_4)\frac{q_1^{\alpha}q_2^{\beta}q_3^{\alpha'}q_4^{\beta'}}{u-m_\pi^2}F_{\pi\gamma\gamma}(q_2,q_3)\right]\,,
\end{align}
\end{widetext}
where $s=(q_1+q_2)^2$, $t=(q_1-q_3)^2$ and $u=(q_1-q_4)^2$ and we have used the shorthand notation $v_i(z)\equiv v(z,Q_i)$. The tensors $T^{\mu}_{ij}$ are defined as
\begin{align}
T^{\mu}_{ij}(z)=\bigg[q_{i}^{\mu}v_i(z)\partial_{z}v_j(z)-q_{j}^{\mu}v_j(z)\partial_{z}v_i(z)\bigg]\,.  
\end{align}
The closed expression of eq.~(\ref{VVVVM}) for the hadronic light-by-light tensor is all that is needed for the evaluation of the contribution to the anomalous magnetic moment. However, one of the virtues of having a consistent model with analytical control is that a number of issues can be examined in detail. This we will do in the following sections.

\section{Longitudinal piece and pion-exchange dominance}\label{sec:4}

Based on large-$N_c$ arguments and dimensional power counting, there is agreement that the pion exchange contribution is the dominant piece of the HLbL. 

\begin{figure}[t]
\begin{center}
\includegraphics[width=0.48\textwidth]{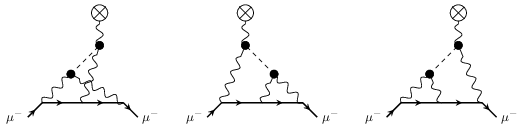}
\end{center}
\caption{The three one-pion exchange HLbL diagrams. 
The dashed line denotes the pion propagator and the black dots the $F_{\pi\gamma\gamma}$ form factors. Photon momenta assignments are the same as in fig.\ref{fig:1}, i.e., they point inwards.} \label{fig:3}
\end{figure}

Equation~(\ref{VVVVM}) contains, as expected, the pion contribution to the hadronic light-by-light tensor as the product of the $\pi\gamma\gamma$ form factors connected by a pion propagator (see fig.~\ref{fig:3}). In turn, the first term on each line accounts for the contribution of the full tower of axial-vector states. 

In order to understand better the structure of the HLbL tensor, it is convenient to project out its longitudinal and transverse parts, which can be done using eq.~(\ref{decomp1}).

The longitudinal component, which also contains the pion contribution, can be expressed in terms of three tensorial structures,
\begin{align}\label{long}
\Pi^{\mu\nu\lambda\rho}_{\parallel}(q_j)={\cal{W}}^{\parallel}_{12;34}T^{(1)}_{\mu\nu\lambda\rho}+{\cal{W}}^{\parallel}_{13;24}T^{(2)}_{\mu\nu\lambda\rho}+{\cal{W}}^{\parallel}_{14;23}T^{(3)}_{\mu\nu\lambda\rho}\,,
\end{align}
where
\begin{align}
T^{(1)}_{\mu\nu\lambda\rho}=\varepsilon_{\mu\nu\alpha\beta}\varepsilon_{\lambda\rho\alpha'\beta'} q_1^{\alpha}q_2^{\beta}q_3^{\alpha'}q_4^{\beta'}
\end{align}
and the tensors $T^{(2)}_{\mu\nu\lambda\sigma}$ and $T^{(3)}_{\mu\nu\lambda\sigma}$ are the crossed-symmetric ones. Defining, as in \cite{Colangelo:2015ama}, the crossing operations $\mathcal{C}_{14}=\{q_1\leftrightarrow q_4,\mu\leftrightarrow \sigma\}$ and $\,\mathcal{C}_{13}=\{q_1\leftrightarrow q_3,\mu\leftrightarrow \lambda\}$, they are related by 
\begin{align}
T^{(2)}_{\mu\nu\lambda\sigma}=\mathcal{C}_{14}~ T^{(1)}_{\mu\nu\lambda\sigma}; \qquad T^{(3)}_{\mu\nu\lambda\sigma}=\mathcal{C}_{13}~ T^{(1)}_{\mu\nu\lambda\sigma}\,.
\end{align}
The longitudinal form factors can be shown to take the simplified form 
\begin{align}
{\cal{W}}^{\parallel}_{12;34}&(q_j;m_{\pi}^2)=F_{\pi\gamma\gamma}(q_1,q_2)\frac{1}{s-m_\pi^2}F_{\pi\gamma\gamma}(q_3,q_4)\nonumber\\
&\!\!\!\!\!\!\!\!\!\!\!\!\!\!\!-\frac{2c^2}{\lambda}\int dz \int dz' v_1(z)v_2(z)\frac{\partial_z\partial_{z'}G_A^{\parallel}(z,z')}{s}v_3(z')v_4(z')\,,
\end{align}
with similar expressions for ${\cal{W}}^{\parallel}_{13;24}$ and ${\cal{W}}^{\parallel}_{14;23}$. The second line above can be easily obtained from the first term in eq.~(\ref{VVVVM}) by using the antisymmetry of the Levi-Civita tensors and integration by parts. 

Using the expression (see Appendix \ref{app:1})
\begin{align}
\partial_z\partial_{z'}G_A^{\parallel}(z,z')=\frac{z_0^2}{2}\bigg[\alpha'(z)\alpha'(z')+\alpha'(z)\delta(z-z')\bigg]\,,
\end{align}
the longitudinal form factor takes the form
\begin{align}\label{long1}
{\cal{W}}^{\parallel}_{12;34}(q_j;m_{\pi}^2)&=F_{\pi\gamma\gamma}(q_1,q_2)\frac{1}{s-m_\pi^2}F_{\pi\gamma\gamma}(q_3,q_4)\nonumber\\
&-\left(\frac{2c}{f_{\pi}}\right)^2\frac{1}{s}\int dz \alpha'(z)v_1(z)v_2(z)v_3(z)v_4(z)\nonumber\\
&-F_{\pi\gamma\gamma}(q_1,q_2)\frac{1}{s}F_{\pi\gamma\gamma}(q_3,q_4)\,,
\end{align} 
where we have used that $z_0^2=8\lambda f_{\pi}^{-2}$ and the definition of the pion transition form factor in eq.~(\ref{pigammagammaff}).

The previous expression shows explicitly that the contribution of the whole tower of axial-vector states consists of a factorizable and a nonfactorizable piece in five dimensions. In four dimensions, they correspond to a propagating piece and a contact term, respectively. In the chiral limit, one can easily check that the axial-vector propagating piece and the pion contribution cancel each other and one is left with the contact term. Explicitly,
\begin{align}\label{long2}
{\cal{W}}^{\parallel}_{12;34}&(q_j;m_{\pi}^2)=-\left(\frac{2c}{f_{\pi}}\right)^2\frac{1}{s}\int dz \alpha'(z)v_1(z)v_2(z)v_3(z)v_4(z)\nonumber\\
&+F_{\pi\gamma\gamma}(q_1,q_2)\frac{m_{\pi}^2}{s(s-m_\pi^2)}F_{\pi\gamma\gamma}(q_3,q_4)\,.
\end{align}
The structure of this expression and, in particular, the presence of the contact term, is mostly dictated by the chiral anomaly, as we will show more explicitly in the next section. 

At very low energies (still in the chiral limit) the integral appearing in the contact term can be easily evaluated. Using that $v_j(z,0)=1$ and the explicit expression for $\alpha(z)$, one finds
\begin{align}
\lim_{s\rightarrow 0}{\cal{W}}^{\parallel}_{12;34}(0;m^2_\pi=0)&=\left(\frac{2c}{f_{\pi}}\right)^2\frac{1}{s}\equiv \frac{F^2_{\pi\gamma\gamma}(0,0)}{s}\,.
\end{align}
The result is actually the same that one would have obtained by dropping the axial-vector tower and considering only the pion exchange contribution. This is of course not a coincidence. At very low energies only the pion is a dynamical degree of freedom and it is entirely responsible for fulfilling the chiral anomaly. This is the content of the Wess-Zumino-Witten term in chiral perturbation theory, which our model also contains, and actually fixes the value of $F_{\pi\gamma\gamma}(0,0)$, as we have already shown. It is clear that, at higher energies, anomaly matching requires the participation of resonance states other than the pion. However, since no first-principle description of the strong interactions exists in the intermediate energy regime, it is not known how this is implemented in detail. 

The result in eq.~(\ref{long2}) is precisely the way the model implements the chiral anomaly in a consistent way {\it{at all energy scales}}. The expression for the resummed axial-vector contributions and the precise cancellation of the pion contribution in the chiral limit can be therefore seen in this light as a sort of sum rule to enforce anomaly matching at all energies. This interpretation will be reinforced in the following section, where we will look into the chiral anomaly in a more explicit fashion. 

\section{Anomaly matching in the $VVA$ correlator}\label{sec:5}

The best way to uncover the role of the axial anomaly in the HLbL tensor is to consider the three-point correlator
\begin{align}\label{VVA}
\Gamma_{\mu\nu\lambda}(q_3,q_4)&=i\int d^4xd^4y~e^{-i(q_3\cdot x+q_4\cdot y)}\nonumber\\
&\times \langle 0|T\left\{j_{\mu}^{\mathrm{em}}(x)j_{\nu}^{\mathrm{em}}(y)j^5_{\lambda}(0)\right\}|0\rangle\,.
\end{align}
This correlator has been studied in great detail in, e.g., \cite{Vainshtein:2002nv,Knecht:2003xy}. In \cite{Knecht:2003xy} it was shown that, on general grounds, $\Gamma_{\mu\nu\lambda}$ can be decomposed into four independent tensorial structures: one longitudinal and three transverse, which are associated with four different scalar functions. 

\begin{figure}[t]
\begin{center}
\includegraphics[width=0.4\textwidth]{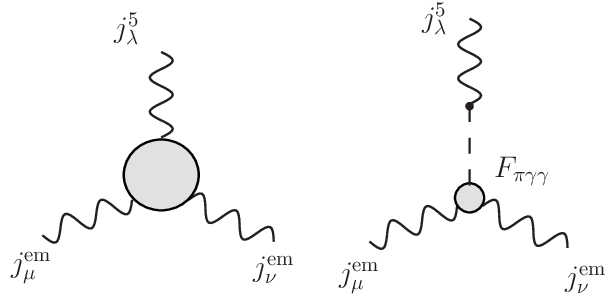}
\end{center}
\caption{Diagrams contributing to the $VVA$ correlator. The one on the left is the axial-vector contribution, where the blob contains the nontrivial momentum dependence of eqs.~(\ref{sep1}) and (\ref{sep2}). The diagram on the right is the pion exchange contribution.}\label{fig:4}
\end{figure}

In the holographic model, the $VVA$ correlator can be computed from the variation of the effective action in eq.~(\ref{effaction}) with respect to the sources. Rewriting the action in terms of the transverse and longitudinal axial-vector sources, integrating by parts and using the boundary conditions for the vector and axial-vector fields, the result is
\begin{align}
(S_{{\mathrm{CS}}}^{(3)})^{\perp}&=\frac{2c}{3}\varepsilon^{\mu\nu\lambda\rho}\int d^4x\,{\hat{a}}_{\mu}^{\perp}(x)\partial_{\nu}{\hat{v}}_{\lambda}(x){\hat{v}}_{\rho}(x)\nonumber\\
&\left[1+3\int_{0}^{z_0} dz\, a(x,z)v(x,z)v'(x,z)\right]\label{sep1}\,,\\
(S_{{\mathrm{CS}}}^{(3)})^{\parallel}&=\frac{c}{3}\varepsilon^{\mu\nu\lambda\rho}\int d^4x\,\frac{\partial^{\alpha}{\hat{a}}_{\alpha}^{\parallel}(x)}{\Box}\partial_{\nu}{\hat{v}}_{\lambda}(x)\partial_{\mu}{\hat{v}}_{\rho}(x)\nonumber\\
&\left[1+3\int_0^{z_0} dz\, \alpha'(z)v(x,z)v(x,z)\right]\label{sep2}\,,
\end{align}
where the first structure in each equation is a boundary term, while the second one has a nontrivial profile and energy dependence. Apart from a local contribution, the correlator also contains a pion-exchange contribution (see fig.~\ref{fig:4}). Explicitly, the pion contribution is generated from the effective action
\begin{align}
&S_{\rm{eff}}^{(\pi)}=\int d^4x \left[\frac{1}{2}\partial_{\mu}\pi(x)\partial^{\mu}\pi(x) + f_{\pi}{\hat{a}}_{\mu}^{\parallel}(x)\partial^{\mu}\pi(x)\right.\nonumber\\
&\left.+\frac{c}{f_{\pi}}\varepsilon_{\mu\nu\lambda\rho}\pi(x)\int_0^{z_0}dz\,\alpha'(z)\partial_{\mu}V_{\nu}(x,z)\partial_{\lambda}V_{\rho}(x,z)\right]\,,
\end{align}
where the first line comes entirely from eq.~(\ref{Seff2}) by using the decomposition of eq.~(\ref{decomposition}).

From the previous terms in the effective action it is straightforward to obtain the expression for $\Gamma_{\mu\nu\lambda}$. The longitudinal part of the correlator yields
\begin{align}
\Gamma_{\mu\nu\lambda}^{\parallel}(q_3,q_4)&=t^{\parallel}_{\mu\nu\lambda}\left[\frac{2c}{3q_3^2}\left(1+3\frac{F_{\pi\gamma\gamma}(q_3,q_4)}{F_{\pi\gamma\gamma}(0,0)}\right)\right.\nonumber\\
&\left.-\frac{2c}{q_3^2-m_{\pi}^2}\frac{F_{\pi\gamma\gamma}(q_3,q_4)}{F_{\pi\gamma\gamma}(0,0)}\right]\label{longVVA}\,,
\end{align}
where $t^{\parallel}_{\mu\nu\lambda}$ is the longitudinal tensor
\begin{equation}
t^{\parallel}_{\mu\nu\lambda}=q_{3\,\lambda}\varepsilon_{\mu\nu\alpha\beta}q_4^{\alpha}q_3^{\beta}\,,
\end{equation}
defined as in \cite{Knecht:2003xy}.

The transverse part is less straightforward to obtain. The reason is that the model we are using necessarily describes the {\it{consistent}} anomaly, which is the only one that can be derived from an action~\cite{Bardeen:1984pm}.  However, in the presence of gauge fields, only the {\it{covariant}} anomaly is compatible with the Ward identities. This change of prescription can be interpreted as a different definition of the chronological $T$ product. Therefore, with the effective action of eq.~(\ref{effaction}) one is not computing the correlator defined in eq.~(\ref{VVA}), but instead 
\begin{align}
{\hat{\Gamma}}_{\mu\nu\lambda}(q_1,q_2)&=i\int d^4xd^4y~e^{i(q_1\cdot x+q_2\cdot y)}\nonumber\\
&\times \langle 0|{\hat{T}}\left\{j_{\mu}^{\mathrm{em}}(x)j_{\nu}^{\mathrm{em}}(y)j^5_{\lambda}(0)\right\}|0\rangle\,,
\end{align}
where ${\hat{T}}$ is associated with the consistent anomaly. In this prescription, the photon contains both transverse and longitudinal components. Taking this into account, one can check that the Ward identities are:
\begin{align}
q_3^{\mu}{\hat{\Gamma}}_{\mu\nu\lambda}&=\frac{N_c}{12\pi^2}\varepsilon_{\alpha\beta\nu\lambda}q_4^{\alpha}q_3^{\beta}\,,\nonumber\\
q_4^{\nu}{\hat{\Gamma}}_{\mu\nu\lambda}&=-\frac{N_c}{12\pi^2}\varepsilon_{\alpha\beta\mu\lambda}q_4^{\alpha}q_3^{\beta}\,,\nonumber\\
(q_3+q_4)^{\lambda}{\hat{\Gamma}}_{\mu\nu\lambda}&=-\frac{N_c}{12\pi^2}\varepsilon_{\alpha\beta\mu\nu}q_4^{\alpha}q_3^{\beta}\,,
\end{align}
which are indeed the ones corresponding to the consistent anomaly.

The relation between both prescriptions involves a Bardeen-Zumino polynomial of the form~\cite{Bardeen:1984pm}
\begin{align}
\Gamma_{\mu\nu\lambda}={\hat{\Gamma}}_{\mu\nu\lambda}+\frac{N_c}{12\pi^2}\varepsilon_{\mu\nu\lambda\alpha}(q_1-q_2)^{\alpha}\label{eq:BZ1}\,,
\end{align}
such that in the covariant prescription one recovers the well-known Adler-Bardeen results
\begin{align}
q_3^{\mu}{{\Gamma}}_{\mu\nu\lambda}&=0\,,\nonumber\\
q_4^{\nu}{{\Gamma}}_{\mu\nu\lambda}&=0\,,\nonumber\\
(q_3+q_4)^{\lambda}{{\Gamma}}_{\mu\nu\lambda}&=-\frac{N_c}{4\pi^2}\varepsilon_{\alpha\beta\mu\nu}q_4^{\alpha}q_3^{\beta}\,.
\label{eq:BZ}
\end{align}

In the particular kinematic configuration where $q_4$ and $(q_3+q_4)^2$ go to zero, the four independent tensorial structures reduce to just two \cite{Czarnecki:2002nt,Knecht:2003xy},
\begin{align}
t^{\parallel}_{\mu\nu\lambda}&=q_3^{\lambda}\varepsilon_{\mu\nu\alpha\beta}q_4^{\alpha}q_3^{\beta}\,, \nonumber\\
t^{\perp}_{\mu\nu\lambda}&=q_3^2\varepsilon_{\mu\nu\lambda\rho}q_4^{\rho}-q_3^{\nu}\varepsilon_{\mu\rho\alpha\beta}q_4^{\alpha}q_3^{\beta}\nonumber \\
& \;\;-q_3^{\lambda}\varepsilon_{\mu\nu\alpha\beta}q_4^{\alpha}q_3^{\beta} + \mathcal{O}((q_3+q_4)^2)\,.
\end{align}
Accordingly, in this limit there are only two independent kinematic invariants, defined as
\begin{align}
\Gamma_{\mu\nu\lambda}(q_3,q_4)=\frac{1}{24\pi^2}\left[\omega_L(q_3) t^{\parallel}_{\mu\nu\lambda} +\omega_T(q_3) t^{\perp}_{\mu\nu\lambda}\right]\,.
\end{align}
The longitudinal function $\omega_L$ is known to be fixed by the anomaly to
\begin{align}
\omega_L(q_3)=-\frac{2N_c}{q_3^2}\,.
\end{align}

This result is exact in the chiral limit, and gets corrections only from nonperturbative contributions~\cite{Vainshtein:2002nv}. $\omega_T$ instead depends on the dynamics of axial-vector exchange. Both functions are, however, linked at very high energies, where they satisfy the well-known expression~\cite{Vainshtein:2002nv}
\begin{align}\label{OPE}
\lim_{Q_3\to \infty}\bigg[\omega_L(Q_3)-2\omega_T(Q_3)\bigg]=0\,.
\end{align}

It is relatively easy to check that the previous general QCD results get reproduced with our model. If one considers eq.~(\ref{longVVA}) in the chiral limit, one finds a cancellation between the pion contribution and the energy-dependent part of the longitudinal vertex. This is the same cancellation that we already discussed in the previous section. In the chiral limit, therefore, the longitudinal part is saturated by the boundary term. Notice that this cancellation between the whole tower of axial-vector states and the pion, which is naturally implemented by the model, is the only way to have 
\begin{align}
\omega_L(q_3)\sim \frac{1}{q_3^2}
\end{align}
at all energies. The cancellation of the pion contribution against a collective effect of the whole axial-vector tower (in the chiral limit) is thus a sum rule enforced by the chiral anomaly. 

As constructed in eq.~(\ref{eq:BZ1}), the Bardeen-Zumino term cancels the boundary terms in the transverse part and brings the longitudinal part to the Adler-Bardeen value. In this prescription, the predictions for $\omega_L$ and $\omega_T$ in Euclidean space at $\mathcal{O}(Q_4^2)$ read 
\begin{align}\label{structure}
\omega_L(Q_3)&=\frac{2N_c}{Q_3^2}-\left(\frac{2N_c}{Q_3^2}-\frac{2N_c}{Q_3^2+m_{\pi}^2}\right)\frac{F_{\pi\gamma\gamma}(Q_3,0)}{F_{\pi\gamma\gamma}(0,0)}\,,\nonumber\\
\omega_T(Q_3)&=-\frac{2N_c}{Q_3^2}\int_0^{z_0}dz a(z,Q_3)v_3'(z,Q_3)\,.
\end{align}

The integral in $\omega_T$ can actually be solved analytically and its expression considerably simplified. In order to do so, it is convenient to rewrite~\cite{Son:2010vc} 
\begin{align}
a(z,q_3)v_3'(z,q_3)&=\frac{1}{2}\partial_{z}\big[a(z,q_3)v_3(z,q_3)\big]\nonumber\\
&\!\!\!\!\!\!\!\!\!\!\!\!\!\!\!\!\!\!\!\!+\frac{1}{2}\big[a(z,q_3)v_3'(z,q_3)-a'(z,q_3)v_3(z,q_3)\big]\,.
\end{align}
The first piece is a boundary term while the second one can be shown to be linear in $z$ (see eq.~(\ref{Wronskian})). The result for both form factors can therefore be written in the rather compact form:
\begin{align}
\omega_L(Q_3)&=\frac{2N_c}{Q_3^2}\left[1-\frac{m_{\pi}^2}{Q_3^2+m_{\pi}^2}\frac{F_{\pi\gamma\gamma}(Q_3,0)}{F_{\pi\gamma\gamma}(0,0)}\right] + \mathcal{O}(Q_4^2)\,,\nonumber\\
\omega_T(Q_3)&=\frac{N_c}{Q_3^2}-\frac{N_c}{2}z_0^2(\xi_0+\xi_1) + \mathcal{O}(Q_4^2)\,.
\end{align}
As expected, in the chiral limit the pion contribution gets cancelled by part of the axial-vector one such that $\omega_L$ is structureless, as already noticed before. Corrections to this expression are proportional to the pion mass and are therefore of nonperturbative nature, in compliance with QCD. We stress that this is a consistency check that the anomaly is correctly implemented in the model. The study of the $VVA$ correlator thus reveals that the nonfactorizable piece in the longitudinal part of the HLbL tensor that we observed in the previous section has the same origin as the contact term in the $VVA$ correlator that reproduces the Adler-Bardeen result for the chiral anomaly.

\begin{figure}[t]
\begin{center}
\includegraphics[width=0.4\textwidth]{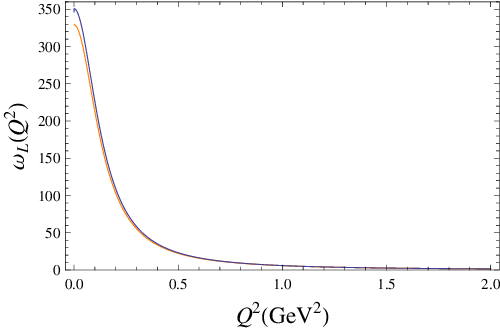}
\vskip 0.2cm
\includegraphics[width=0.4\textwidth]{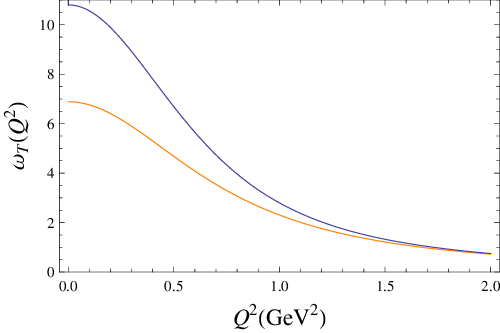}
\caption{Comparison between our predictions for $\omega_{L,T}(Q^2)$ (blue lines) and the ones from ref.~\cite{Melnikov:2003xd} (orange lines). The vertical axis is in units of GeV$^{-2}$.}\label{fig:5}
\end{center}
\end{figure}

Expanding the previous expressions for large momenta, one finds 
\begin{align}
\lim_{Q_3\to \infty}\omega_L(Q_3)&=\frac{2N_c}{Q_3^2}+{\cal{O}}\left(\frac{m_{\pi}^2}{Q_3^6},Q_4^2\right)\,,\nonumber\\
\lim_{Q_3\to \infty}\omega_T(Q_3)&=\frac{N_c}{Q_3^2}+{\cal{O}}(e^{-Q_3z_0},Q_4^2)\,,
\end{align}
which implies that
\begin{align}
\lim_{Q_3\to \infty}\bigg[\omega_L(Q_3)-2\omega_T(Q_3)\bigg]\sim {\cal{O}}\left(\frac{m_{\pi}^2}{Q_3^6},Q_4^2\right)\,.
\end{align} 
This scaling is consistent with the OPE. However, as already emphasized, conformal symmetry in the five-dimensional model ensures that the OPE relations will be satisfied to leading order, but the effects of OPE condensates will in general be missed. Therefore, the expression above has the value of a consistency check rather than a prediction.

Another important observation is that the pion contribution at low energies happens to be 
\begin{align}
\lim_{{\hat{q}}\to 0}\omega_L^{(\pi)}(q_3)=-\frac{2N_c}{q_3^2}\,,
\end{align} 
and one would be tempted to conclude that the pion saturates the anomaly at low energies. This interpretation is not wrong but it is misleading: in the $m_{\pi}\to 0$ limit, the pion and the dynamical axial-vector contribution cancel analytically at all energies. Based on the previous derivation, one is forced to conclude that the pion does not saturate $\omega_L$, although it is fundamental to make sure that the result is consistent with the anomaly. 

\begin{figure}[t]
\begin{center}
\includegraphics[width=0.4\textwidth]{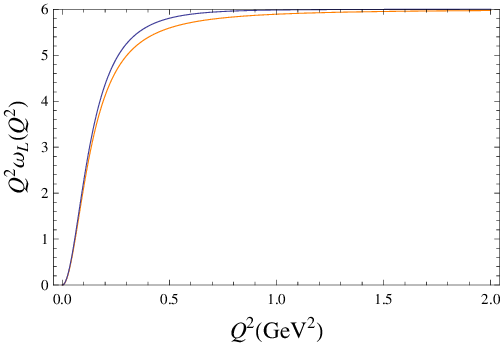}
\vskip 0.2cm
\includegraphics[width=0.4\textwidth]{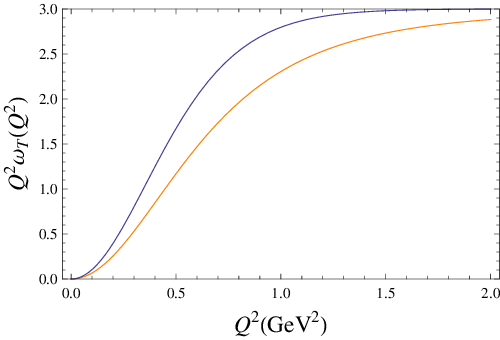}
\end{center}
\caption{$Q^2\omega_{L,T}(Q^2)$ for both our model and the one of ref.~\cite{Melnikov:2003xd}. The asymptotic limit agrees in all cases with the OPE results.}\label{fig:6}
\end{figure}

The form of the chiral corrections to both form factors and its behavior at intermediate energies are not known from first principles and are therefore a prediction of the model. At low energies both form factors go to a constant, namely
\begin{align}
\lim_{Q_3\to 0}\omega_L(Q_3)&=\frac{2N_c}{m_{\pi}^2}(1+a_{\pi})+{\cal{O}}(Q_3^2,Q_4^2)\,,\nonumber\\
\lim_{Q_3\to 0}\omega_T(Q_3)&=\frac{3N_c}{8}z_0^2+{\cal{O}}(Q_3^2,Q_4^2)\,,
\end{align}
where $a_{\pi}$ is the slope of $F_{\pi\gamma\gamma}$ at zero momentum (see eq.~(\ref{slope})). The predictions of the present model can be compared, for instance, with the expressions used in~\cite{Melnikov:2003xd}:
\begin{align}
\big[\omega_L(Q_3)\big]^{\rm{MV}}&=\frac{2N_c}{Q_3^2+m_{\pi}^2}\,,\nonumber\\
\big[\omega_T(Q_3)\big]^{\rm{MV}}&=\frac{N_c}{m_{a_1}^2-m_{\rho}^2}\left[\frac{m_{a_1}^2-m_{\pi}^2}{Q_3^2+m_{\rho}^2}-\frac{m_{\rho}^2-m_{\pi}^2}{Q_3^2+m_{a_1}^2}\right]\,.
\end{align} 
The results are illustrated in figs.~\ref{fig:5} and \ref{fig:6}, where one can see that the main difference between both models mostly affect $\omega_T$.  

\section{The Melnikov-Vainshtein limit}\label{sec:6}

The closed expression that we obtained for the light-by-light tensor in eq.~(\ref{VVVVM}) can be extrapolated to large Euclidean momenta and compared with the different short-distance constraints derived from the OPE of QCD. This is a necessary consistency check before the numerical analysis of the next section. The more constraints the expression satisfies, the more reliable the predictions will be. Conversely, if a constraint is not satisfied, one can estimate its impact by comparing with models that do implement it.

Relevant for the evaluation of the $(g-2)_{\mu}$ are the limits when $q_4^2=0$. A nontrivial check is to find out how the HLbL behaves when all virtual photons have large momenta, $Q_1^2=Q_2^2=Q_3^2\equiv Q^2\gg \Lambda_{\rm{QCD}}^2$. For the longitudinal component of the HLbL tensor, the quark loop diagram in QCD gives~\cite{Melnikov:2003xd} 
\begin{align}
{\cal{W}}_{12;34}^{\parallel}=-\frac{4}{9\pi^2Q^4}\sim -\frac{0.44}{\pi^2Q^4}\,.
\end{align}
One can find the corresponding expression in the model by applying the high-energy limit to eq.~(\ref{long1}). The first thing to note is that this limit cannot be fulfilled by the pion contribution, which falls off like $Q^{-6}$. The relevant piece comes instead from the axial-vector tower, from which one obtains
\begin{align}
{\cal{W}}_{12;34}^{\parallel}=-\frac{N_c}{3\pi^2Q^4} \int_0^\infty x^4 K_1(x)^3\; dx\sim -\frac{0.36}{\pi^2Q^4}\,. 
\end{align}
Numerically, this amounts to $80\%$ of the OPE coefficient. This deficit is not necessarily a mismatch, given that other hadronic contributions to the HLbL not included in our model (e.g., pseudoscalar mesons) are also expected to contribute to the quark loop matching.
  
Particularly interesting are the limits where only two photons have large virtualities. Without loss of generality one can consider $Q_1^2\simeq Q_2^2\gg Q_3^2\gg \Lambda_{QCD}^2$, with the remaining two possibilities generated by crossing symmetry. In the context of the $(g-2)_{\mu}$, these limits were first explored in \cite{Melnikov:2003xd}, where the relation with the anomaly was emphasized. 

The key object to study this limit is the product of two electromagnetic currents:
\begin{align}
W^{\mu\nu}(q_1,q_2)=\int d^4x\int d^4y\, e^{i(q_1\cdot x+q_2\cdot y)} T\big\{j^{\mu}_{\rm{em}}(x),j^{\nu}_{\rm{em}}(y)\big\}
\end{align}
in the kinematical limit
\begin{align}\label{MVkinematic}
Q_1=\xi Q-\frac{Q_3}{2}\,;\qquad Q_2=-\xi Q-\frac{Q_3}{2}\,,
\end{align}
where $\xi$ is large and all momenta are spacelike. In this limit, the OPE gives 
\begin{align}
\lim_{\xi\to\infty}W^{\mu\nu}&\left(\xi Q-\frac{Q_3}{2},-\xi Q-\frac{Q_3}{2}\right)=\frac{1}{\xi}\frac{2i}{Q^2}\epsilon^{\mu\nu\lambda\rho}Q_{\lambda}\nonumber\\
&\qquad\quad\times\sum_{a}{\hat{d}}^{a\gamma\gamma}\int d^4z e^{-i q_3\cdot z}j_{5\rho}^{(a)}(z)\,,
\end{align}
with $j_{5\rho}^{(a)}(z)={\bar{q}}{\hat{Q}}^2\gamma_{\rho}\gamma_5 q$. This limit shows that a number of the short-distance constraints relevant for the evaluation of the HLbL are actually determined by the axial anomaly through the $VVA$ correlator discussed in the previous section.

Within the model, it is rather straightforward to check that, when the limit $Q_1^2\simeq Q_2^2\gg Q_3^2\gg \Lambda_{QCD}^2$ is taken, the leading term comes from the first line of eq.~(\ref{VVVVM}), i.e. from figs.~2(a) and 2(b). The crossed diagrams are subleading, as expected. 

The pion contribution can be computed using the asymptotic expression in eq.~(\ref{OPEpion}), from which one concludes that it falls off with higher powers of $Q^2$ than the previous OPE demands. The contribution of the axial-vector tower is the relevant piece. For its evaluation it is convenient to consider the integral
\begin{align}
J^{\sigma\alpha\beta}(z,q_1,q_2)&=\int dz^{\prime}G_A^{\alpha\beta}(z,z^{\prime};s)T_{12}^{\sigma}\left(z^{\prime}\right)\,.
\end{align}    
In order to separately analyze the longitudinal and transverse components, we will define
\begin{align}
J^{\sigma\alpha\beta}(z,q_1,q_2)=P_{\perp}^{\alpha\beta}(q_3)J^{\sigma}_{\perp}(z,q_1,q_2)+P_{\parallel}^{\alpha\beta}(q_3)J^{\sigma}_{\parallel}(z,q_1,q_2)\,.
\end{align}
In the limit $Q\gg Q_3$, one can easily show that
\begin{align}
J^{\sigma}_{\perp}(z,Q,Q_3)&=-\frac{2Q^{\sigma}}{Q^2}a(z,Q_3)\left[\frac{1}{3}+\frac{1}{5}\left(\frac{Q_3}{Q}\right)^2+\cdots\right]\,,
\end{align}
while
\begin{align}
J^{\sigma}_{\parallel}(z,Q)=-\frac{2Q^{\sigma}}{3Q^2}\alpha(z)\,.
\end{align} 
The fact that the longitudinal piece is exact to all orders in $Q_3$ is the manifestation of the chiral anomaly, as we will show below.

For the soft momenta, using that 
\begin{align}
\lim_{q\to 0} v(z,q)=1+{\cal{O}}(q^2)f(z)\,,
\end{align}
one easily concludes that
\begin{align}
\lim_{q_4\to 0}T^{\mu}_{34}(z)&=-q_4^{\mu}\partial_{z}v_3(z)+{\cal{O}}(q_4^2)\,.
\end{align}

Plugging the previous expressions back into the HLbL tensor, one finds
\begin{widetext}
\begin{align}\label{VVV}
\Pi_{\mu\nu\lambda\rho}(Q,Q_3)&=\varepsilon_{\mu\nu\alpha\beta}\varepsilon_{\lambda\rho\alpha'\beta'}q_4^{\beta'}\left[-\frac{2c^2}{\lambda}\int dz v_3'(z)\left[P_{\perp}^{\alpha\alpha'}J^{\beta}_{\perp}(z,Q,Q_3)+P_{\parallel}^{\alpha\alpha'}J^{\beta}_{\parallel}(z,Q,Q_3)\right]+\frac{2}{3}\frac{f_{\pi}}{Q^2}\frac{q_1^{\alpha}q_2^{\beta}q_3^{\alpha'}}{Q_3^2+m_{\pi}^2}F_{\pi\gamma\gamma}(Q_3,0)\right]\nonumber\\
&=\varepsilon_{\mu\nu\alpha\beta}\varepsilon_{\lambda\rho\alpha'\beta'}q_4^{\beta'}Q^{\beta}\frac{Q_3^2}{Q^2}\frac{1}{36\pi^2}\left[P_{\perp}^{\alpha\alpha'}\omega_T+P_{\parallel}^{\alpha\alpha'}\omega_L\right]\,,
\end{align}
\end{widetext}
where in the second line we have identified $\omega_L$ and $\omega_T$ using their expressions in eq.~(\ref{structure}). The previous equation shows explicitly that the cancellation between the pion and longitudinal axial-vector contributions that we have observed before is associated with the right expression for $\omega_L$ or, analogously, the correct implementation of the chiral anomaly. Notice that the pion contribution alone, i.e.,
\begin{align}
\omega_L^{(\pi)}(Q_3)\sim \frac{2N_c}{Q_3^2+m_{\pi}^2}F_{\pi\gamma\gamma}(Q_3,0)\,,
\end{align}
is clearly incompatible with $\omega\sim Q_3^{-2}$ in the chiral limit. The problem is the structure of the form factor, which depends on $Q_3$. For this same reason, it is clear that {\it{no}} single particle exchange can saturate $\omega_L$. In order to satisfy the constraint, an infinite number of (axial-vector) particles is needed. This is precisely what the contact term is indicating. 

We note that the mechanism to saturate the MV constraint with the whole tower of axial-vector states found above follows from imposing the correct implementation of the chiral anomaly. It is therefore not a peculiar feature of our model but a rather generic one. In the next section we will show that this has a substantial impact on the HLbL, thus confirming the numerical importance of the MV constraint. 

\section{Numerical analysis}\label{sec:7}

In eq.~(\ref{VVVVM}) we already wrote down the electromagnetic HLbL tensor in a closed form within our model. In order to perform our numerical analysis and be able to compare with other studies, we will employ standard model-independent techniques.

Quite generically, the HLbL tensor can be expanded as a sum over gauge-invariant Lorentz tensor structures. Out of the 138 structures \cite{Leo:1975fb}, once Ward identities are imposed, one ends up with $43$ kinematic structures. Here we will adopt the formalism introduced in \cite{Colangelo:2015ama}, which builds on previous works \cite{Bardeen:1969aw,Tarrach:1975tu}, where the number of kinematical structures is extended to ensure that they are free of poles and zeros. We will therefore write
\begin{align}
	\label{HLbLBTT}
	\Pi^{\mu\nu\lambda\sigma}(q_1,q_2,q_3) &= \sum_{i=1}^{54} T_i^{\mu\nu\lambda\sigma} \Pi_i(q_1^2,q_2^2,q_3^2)\,, 
\end{align}
where the definitions of the tensors $T_i^{\mu\nu\lambda\sigma}$ can be found in \cite{Colangelo:2015ama}.

Using projection techniques, the two-loop diagram of fig.~\ref{fig:1} can be related to the muon anomalous magnetic moment as follows:
\begin{widetext}
\begin{align}\label{2LoopDiagr}
a_\mu^{\mathrm{HLbL}} &= - \frac{e^6}{48 m_\mu}  \int \frac{d^4q_1}{(2\pi)^4} \frac{d^4q_2}{(2\pi)^4} \frac{1}{q_1^2 q_2^2 (q_1+q_2)^2} \frac{1}{(p+q_1)^2 - m_\mu^2} \frac{1}{(p-q_2)^2 - m_\mu^2}\nonumber\\
			&\times {\mathrm{Tr}}\left( (\slashed p + m_\mu) [\gamma^\rho,\gamma^\sigma] (\slashed p + m_\mu) \gamma^\mu (\slashed p + \slashed q_1 + m_\mu) \gamma^\lambda (\slashed p - \slashed q_2 + m_\mu) \gamma^\nu \right) \; \left( \frac{\partial}{\partial q_4^\rho} \Pi^{\mu\nu\lambda\sigma}(q_1,q_2,-q_4-q_1-q_2) \right) \bigg|_{q_4=0}\,,
\end{align}
\end{widetext}
where $p$ is the muon momentum. Because of the projection above, only 19 independent linear combinations of the 54 $T_i^{\mu\nu\rho\lambda}$ contribute to $a_\mu^\mathrm{HLbL}$ \cite{Leo:1975fb}. Furthermore, due to the symmetries of the two-loop integral, one needs to evaluate eventually only 12 different integrals containing 12 scalar coefficients ${\bar\Pi}_i(q_1,q_2,q_3)$.

Following the general analysis outlined in \cite{Knecht:2001qf,Colangelo:2015ama}, one can perform five out of the eight integrals above using Gegenbauer polynomials, regardless of the specific form of ${\bar\Pi}_i$. The resulting master formula contains then only three integrals and, in terms of Euclidean momenta, takes the form:
\begin{align}\label{master1}
	a_\mu^\mathrm{HLbL}& = \frac{2 \alpha^3}{3 \pi^2} \int_0^\infty dQ_1 \int_0^\infty dQ_2 \int_{-1}^1 d\tau \sqrt{1-\tau^2} Q_1^3 Q_2^3\nonumber\\
	&\qquad\times\sum_{i=1}^{12} \bar T_i(Q_1,Q_2,\tau) \bar \Pi_i(Q_1,Q_2,\tau)\,,
\end{align}
where $Q_1$ and $Q_2$ are the radial components of the momenta. The hadronic scalar functions $\bar \Pi_i$ are  evaluated for the reduced kinematics
\begin{align}
(q_1^2,q_2^2,q_3^2,q_4^2)=(-Q_1^2,-Q_2^2,- Q_1^2 - 2 Q_1 Q_2 \tau - Q_2^2,0)\,.
\end{align}
The complete list of the integral kernels $\bar T_i(Q_1,Q_2,\tau)$ can be found in Appendix B of~\cite{Colangelo:2017fiz}. 

\subsection{Longitudinal contributions}

Given our previous discussion, it is convenient to split the contributions in longitudinal and transverse parts. One can check that the longitudinal contribution is described by the first two structures in eq.~(\ref{master1}). They correspond to eq.~(\ref{long}), i.e.,
\begin{align}
\Pi^{\mu\nu\lambda\rho}_{\parallel}(q_j)={\cal{W}}^{\parallel}_{12;34}T^{(1)}_{\mu\nu\lambda\rho}+{\cal{W}}^{\parallel}_{13;24}T^{(2)}_{\mu\nu\lambda\rho}+{\cal{W}}^{\parallel}_{14;23}T^{(3)}_{\mu\nu\lambda\rho}\,,
\end{align}
where
\begin{align}
T^{(1)}_{\mu\nu\lambda\rho}=\varepsilon_{\mu\nu\alpha\beta}\varepsilon_{\lambda\rho\alpha'\beta'} q_1^{\alpha}q_2^{\beta}q_3^{\alpha'}q_4^{\beta'}\,,
\end{align}
and similarly for the crossed-symmetric $T^{(2)}_{\mu\nu\lambda\sigma}$ and $T^{(3)}_{\mu\nu\lambda\sigma}$. 

The scalar invariants were simplified in Sec.~\ref{sec:4} to the form
\begin{align}\label{decomp}
{\cal{W}}^{\parallel}_{12;34}(q_j;m_{\pi}^2)&=F_{\pi\gamma\gamma}(q_1,q_2)\frac{1}{s-m_\pi^2}F_{\pi\gamma\gamma}(q_3,q_4)\nonumber\\
&-\left(\frac{2c}{f_{\pi}}\right)^2\frac{1}{s}\int dz \alpha'(z)v_1(z)v_2(z)v_3(z)v_4(z)\nonumber\\
&-F_{\pi\gamma\gamma}(q_1,q_2)\frac{1}{s}F_{\pi\gamma\gamma}(q_3,q_4)\,,
\end{align} 
where the first line accounts for the pion contribution and the remaining two are the resummation of the whole axial-vector tower. 

In Sec.~\ref{sec:4} we showed that our model generates a pion transition form factor which has the correct intercept at zero momentum and the right scaling when one or the two photons have large momenta, namely
\begin{align}\label{conditions}
F_{\pi\gamma\gamma}(0,0)&=-\frac{N_c}{12\pi^2 f_\pi}\,,\nonumber\\
\lim_{Q^2\to \infty}F_{\pi\gamma\gamma}(Q^2,Q^2)&=-\frac{2f_{\pi}}{3Q^2}\,,\nonumber\\
\lim_{Q^2\to \infty}F_{\pi\gamma\gamma}(0,Q^2)&=-\frac{2f_{\pi}}{Q^2}\,.
\end{align}
However, in Sec.~\ref{sec:2} we also emphasized that our model predicts 
\begin{align}
\frac{m_{\rho}}{f_{\pi}}=\gamma_{0,1}\sqrt{\frac{6}{N_c}}\pi\sim 10.7\,, 
\end{align}
which is roughly $30\%$ bigger than the experimental number. This is a well-known shortcoming of the present model, which can be amended with more sophisticated versions of it. Such sophistications are beyond the scope of the present paper. In the following we will make two choices for the parameters $c$, $\lambda$ and $z_0$, which emphasize different energy regimes in the HLbL. This will be used as an estimate of the uncertainty in our determination of $a_{\mu}$.

Since the longitudinal contribution to HLbL is the dominant one and, in particular, the pion form factor plays a prominent role, a reasonable criterium is to choose the parameters such that agreement with $F_{\pi\gamma\gamma}$ is achieved. Based on eqs.~(\ref{conditions}), fixing $N_c$ and $f_{\pi}$ to the physical values thus seems to be the reasonable choice, at the price of overshooting $m_{\rho}$.      

However, experimental information exists also for the slope of the form factor at low momentum, defined as
\begin{align}\label{slope}
\lim_{Q\to 0}F_{\pi\gamma\gamma}(Q,0)&=-\frac{N_c}{12\pi^2 f_\pi}\left[1-a_{\pi}\frac{Q^2}{m_{\pi}^2}+\cdots\right]\,.
\end{align}
Using the low-energy expansion for $v(z,Q)$,
\begin{align}
v(z,Q)=1-Q^2\left[1-2\log\frac{z}{z_0}\right]\frac{z^2}{4}+\cdots\,,
\end{align}
one readily finds \cite{Grigoryan:2008cc,Cappiello:2010uy,Leutgeb:2019zpq}
\begin{align}
a_{\pi}=-m_{\pi}^2\int_0^{z_0}dz\alpha'(z)\left[1-2\log\frac{z}{z_0}\right]\frac{z^2}{4}=0.033\,,
\end{align}
which is in excellent agreement with the current world average, $(a_{\pi})_{\mathrm{exp}}=0.0335(31)$, if one fixes $z_0$ to match the physical $m_{\rho}$. Instead, the result is grossly undershot if one fixes $z_0$ with $f_{\pi}$.

In the following, we will thus consider the following choices of parameters:
\begin{align}\label{choices}
\frac{f_{\pi}}{N_c}&= 31\, {\rm{MeV}};\qquad &m_{\rho}&=776\,{\rm{MeV}}\,,\qquad & ({\rm{\bf{Set\, 1}}})\nonumber\\
f_{\pi}&=93\, {\rm{MeV}};\qquad &N_c&=3\,. \qquad & ({\rm{\bf{Set\, 2}}})
\end{align}
Additionally, we will take as input
\begin{align}
\alpha_{\rm{em}}=\frac{1}{137.036};\quad  m_\mu=105.7\,{\rm{MeV}};\quad m_\pi=135\,{\rm{MeV}}\,.
\end{align}
The first choice of parameters ensures the right behavior of $F_{\pi\gamma\gamma}$ at low energies (intercept and slope). An additional perk of fixing $m_{\rho}$ to its physical value is that the axial-vector multiplet masses fall into the right ballpark (see eq.~(\ref{ratiom})). At high energies, it still correctly reproduces the $Q^2$ fall-off behavior, yet it fails to pin down the right coefficients. The second choice of parameters matches the expected short-distance behavior but gives a poor determination of $m_{\rho}$ and the $F_{\pi\gamma\gamma}$ slope.

\begin{figure}[t]
\begin{center}
\includegraphics[width=0.4\textwidth]{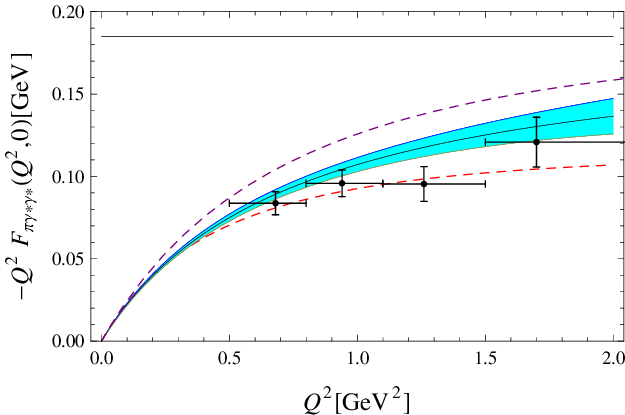}
\end{center}
\caption{$F_{\pi\gamma\gamma}$ with one virtual photon. Data points are taken from~\cite{Behrend:1990sr}. The horizontal line is the asymptotic Brodsky-Lepage limit for large $Q^2$. The lower dashed line is generated with the parameters of {\textbf{Set 1}}, while the upper dashed line is done with {\textbf{Set 2}}. The continuous line is the lattice result of Ref.~\cite{Gerardin:2019vio}, with the corresponding error bands.}\label{fig:7}
\end{figure}

\begin{figure}[t]
\begin{center}
\includegraphics[width=0.4\textwidth]{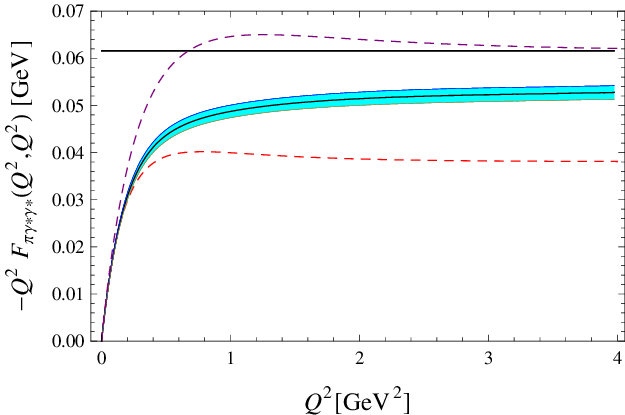}
\end{center}
\caption{$F_{\pi\gamma\gamma}$ when both photons have the same virtuality. The horizontal line is the OPE value given in eq.~(\ref{OPEpion}). Conventions are the same as in fig.~\ref{fig:7}.}\label{fig:8}
\end{figure}

Given the form of the kernels in the expression for $a_{\mu}$, which are peaked in the low-GeV regime, one would be tempted to prioritize the low-energy regime. However, this depends on how fast the asymptotic regime sets in. In figs.~\ref{fig:7} and ~\ref{fig:8} we compare the predictions for the pion form factor with one virtual photon and both photons virtual with equal momenta, respectively, using both sets of parameters in eq.~(\ref{choices}). The lattice result of~\cite{Gerardin:2019vio} is also included for comparison. As expected from our previous discussion, {\bf{Set 1}} fits the experimental data better than {\bf{Set 2}}. However, $a_{\mu}^{\rm{HLbL}}$ is proportional to the integral of the form factor over Euclidean space, and is thus sensitive to global aspects of the form factor. Therefore, there is {\emph{a priori}} no reason to prefer one set of parameters over the other. In the following, we will report our numbers for both sets. Our final number will be the average of them.  
	
\begin{figure}[t]
\begin{center}
\includegraphics[width=0.45\textwidth]{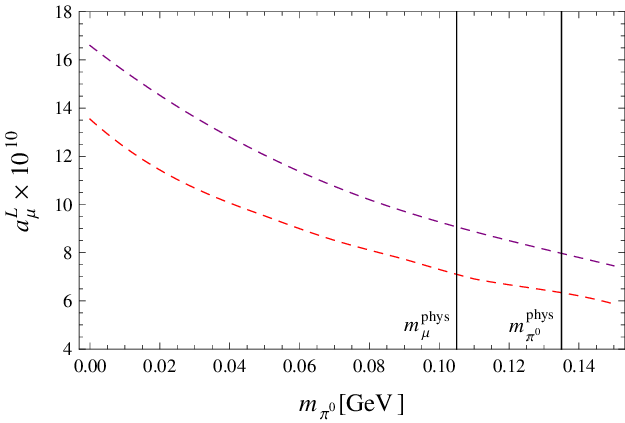}
\end{center}
\caption{The chiral extrapolation of the isovector component of $a_{\mu}^L$. The upper and lower dashed lines correspond to the predictions with {\textbf{Set 2}} and {\textbf{Set 1}}, respectively.}\label{fig:9}
\end{figure}

For the pion contribution we find 
\begin{align}
a_{\mu}^{(\pi)}=(5.7-7.5)\cdot 10^{-10}\,,
\end{align}
where the left and right numbers stand for the {\bf{Set 1}} and {\bf{Set 2}} predictions. This is in excellent agreement with previous determinations, e.g.,
\begin{align}
a_{\mu}^{(\pi)}&=5.7(0.3)\cdot 10^{-10}\,,\qquad {\text{\cite{Hayakawa:1997rq,Hayakawa:2001bb}}}\nonumber\\
a_{\mu}^{(\pi)}&=5.9(0.9)\cdot 10^{-10}\,,\qquad {\text{\cite{Bijnens:1995cc,Bijnens:2001cq}}}\nonumber\\
a_{\mu}^{(\pi)}&=5.8(1.0)\cdot 10^{-10}\,,\qquad {\text{\cite{Knecht:2001qf,Knecht:2001qg}}}\nonumber\\
a_{\mu}^{(\pi)}&=6.8(0.3)\cdot 10^{-10}\,,\qquad {\text{\cite{Greynat:2012ww}}}\nonumber\\
a_{\mu}^{(\pi)}&=6.3(0.3)\cdot 10^{-10}\,.\qquad {\text{\cite{Hoferichter:2018kwz}}}
\end{align}
In turn, the axial-vector contribution to the longitudinal part coming from the isovector $a_1$ and its excitations reads
\begin{align}
a_{\mu}^{(a_1)}&=0.4\cdot 10^{-10}\,.
\end{align} 
for both sets of parameters. The final result for the longitudinal piece in the isovector channel therefore reads 
\begin{align} 
a_{\mu}^L&=(5.7+13.5-13.1)\cdot 10^{-10}=6.1\cdot 10^{-10}\,,\quad  & ({\bf{Set 1}})\nonumber\\
a_{\mu}^L&=(7.5+16.6-16.2)\cdot 10^{-10}=7.9\cdot 10^{-10}\,,\quad  & ({\bf{Set 2}})
\end{align}
where the different contributions are ordered as in eq.~(\ref{decomp}). The second number is the contribution of the contact term, which corresponds to the value for $a_{\mu}^L$ in the chiral limit. Notice that the axial-vector contribution is the result of a large numerical cancellation between the second and third terms. The result above also shows that the chiral corrections to $a_{\mu}^L$ amount to a $50\%$ decrease of its value (see fig.~\ref{fig:9}). 

The contributions of the isoscalar pseudoscalars will be estimated by simply using the physical values for masses and decay constants,
\begin{align}
f_{\eta}&=93\,{\rm{MeV}};\qquad & f_{\eta'}&=74\,{\rm{MeV}}\,,\\
m_{\eta}&=548\,{\rm{MeV}};\qquad & m_{\eta'}&=958\,{\rm{MeV}}\,.
\end{align} 
These numbers are not the ones predicted by the model, which has an exactly massless Goldstone nonet with a common decay constant. The introduction of masses can be argued exactly as we did with the pion: their effects are limited to the pseudoscalar propagators without affecting the dynamics of the model. From the five-dimensional perspective, this can be achieved with the introduction of flavor-dependent boundary terms. For our purposes, we will simply add the different masses by hand. Breaking the degeneracy of the decay constants to fit the experimental values for the $\eta$ and $\eta'$ can also be done as long as one consistently correlates it with the corresponding isoscalar axial-vector channel, such that the sum rules that preserve anomaly matching remain in place. In practice, this can be done by modifying the parameter sets to
\begin{align}
\frac{f_{\eta'}}{N_c}&= 24.7\, {\rm{MeV}};\quad &m_{\rho}&=776\,{\rm{MeV}}\,,\quad &({\rm{\textbf{Set\, 1}}})\nonumber\\
f_{\eta'}&=74\, {\rm{MeV}};\quad &N_c&=3\,, \quad &({\rm{\textbf{Set\, 2}}})
\end{align}
for the $\eta'$ and axial-vector $f_1^*(1420)$ tower, and similarly for the $\eta$ and axial-vector $f_1(1285)$ tower.

Our results are
\begin{align}
a_{\mu}^{(\eta)}&=(1.4-2.1)\cdot 10^{-10};\,\,\, &a_{\mu}^{(\eta')}&=(1.0-1.6)\cdot 10^{-10}\,,\nonumber\\
a_{\mu}^{(f_1)}&=0.4\cdot 10^{-10};\,\,\, &a_{\mu}^{(f_1^*)}&=0.6\cdot 10^{-10}\,.
\end{align}

We emphasize that the axial-vector contributions are the result of resumming full towers of states, with the first state shown as representative. A comparison of these numbers with the contributions of the lowest-lying axial-vector mesons reported in~\cite{Jegerlehner:2017gek,Pauk:2014rta,Danilkin:2019mhd,Roig:2019reh} is therefore not meaningful. We also note that our prescription to satisfy the anomaly implies that the $f_1$ and $f_1^*$ towers of states have the same flavor structure as $\eta$ and $\eta'$, which is not phenomenologically favored. One can improve on the low-energy phenomenology of axial-vector mesons~\cite{Leutgeb:2019gbz} but it is not easy to preserve the anomaly at the same time. Based on these considerations, we place more confidence in our estimate for the total axial-vector contribution rather than on less inclusive, e.g. single-particle, axial-vector contributions.
 
\begin{figure}[t]
\begin{center}
\includegraphics[width=0.4\textwidth]{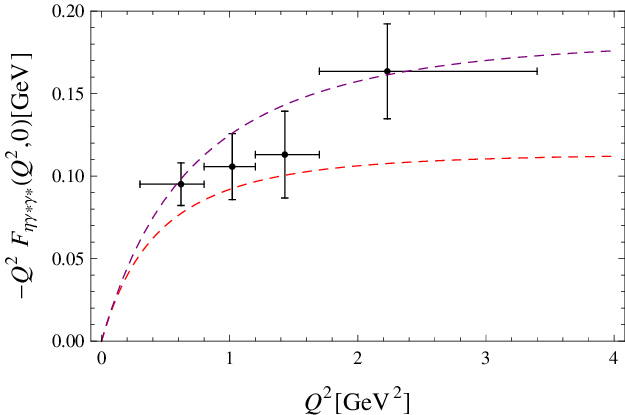}
\includegraphics[width=0.4\textwidth]{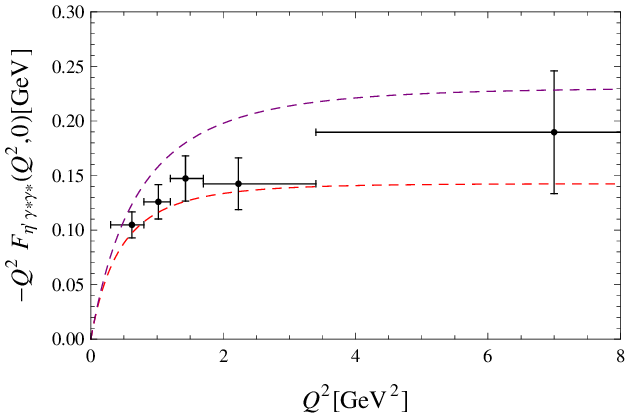}
\end{center}
\caption{$F_{\eta\gamma\gamma}$ and $F_{\eta'\gamma\gamma}$ with one virtual photon. Experimental data are taken from~\cite{Behrend:1990sr}. The lower dashed line is generated with the parameters of {\textbf{Set 1}}, while the upper dashed line is done with {\textbf{Set 2}}.}\label{fig:10}
\end{figure}

Regarding the pseudoscalars, the numbers are comparable with the ones quoted in the literature, e.g.,
\begin{align}
a_{\mu}^{(\eta)}&=1.3(0.1)\cdot 10^{-10};\quad &a_{\mu}^{(\eta')}&=1.2(0.1)\cdot 10^{-10}\,,\quad {\text{\cite{Knecht:2001qf}}}
\end{align}
and the more recent determinations in~\cite{Guevara:2018rhj,Eichmann:2019tjk,Raya:2019dnh}. A comparison of the transition form factors resulting from both parameter sets is provided in fig.~\ref{fig:10}.

Adding all the contributions up, our final number for the longitudinal contribution of axial vectors and Goldstone modes is
\begin{align}
a_{\mu}^{L}=(9.6-12.6)\cdot 10^{-10}\,.
\end{align} 

\begin{table}[t]
\begin{ruledtabular}
\begin{tabular}{ccc}
 & $\mbox{\bf Set 1}$ & $\mbox{\bf Set 2}$ \\
\hline\\
$a_\mu^{\rm{PS}}(\pi^0+\eta+\eta')$ & 8.1 (5.7+1.4+1.0) & 11.2 (7.5+2.1+1.6) \\
\\
$a_\mu^{A_L}(a_1+f_1+f_1^*)$ & 1.4 (0.4+0.4+0.6) & 1.4 (0.4+0.4+0.6) \\
\\
\hline\\
$a_\mu^{L} (a_\mu^{\rm{PS}}+a_\mu^{A_L})$ & 9.6 & 12.6 \\
\\
\hline\\
$a_\mu^{T} (a_1+f_1+f_1^*)$ & 1.4 (0.4+0.4+0.6) & 1.4 (0.4+0.4+0.6) \\
\\
\hline\\
$a_\mu$ & 11.0 & 14.0 \\
\end{tabular}
\end{ruledtabular}
\caption{\label{tab:1} Results for the longitudinal and transverse contributions to $a_{\mu}^{\rm{HLbL}}\;\times\;10^{10}$ for the set of values described in the text. In parenthesis, the separate contributions to each entry. The labels for the axial vectors are understood to take into account not just the lowest-lying states, but also the whole tower of excitations.}
\end{table}
\subsection{Transverse contributions}

The transverse terms of the HLbL tensor collect the remaining part of the axial-vector contributions. They represent only a small correction to the HLbL value, but they are considerably harder to work out. Explicitly, they follow from
\begin{align}\label{masterT}
	a_\mu^\mathrm{HLbL}& = \frac{2 \alpha^3}{3 \pi^2} \int_0^\infty dQ_1 \int_0^\infty dQ_2 \int_{-1}^1 d\tau \sqrt{1-\tau^2} Q_1^3 Q_2^3\nonumber\\
	&\qquad\times\sum_{i=3}^{12} \bar T_i(Q_1,Q_2,\tau) \bar \Pi_i(Q_1,Q_2,\tau)\,,
\end{align}
with the first two scalar factors, which are the longitudinal contributions, subtracted. The expressions for the scalar functions in our model are given in Appendix \ref{app:3}.

The final values one obtains for the whole towers of isovector and isoscalar axial-vector states is 
\begin{align}
a_\mu^{T}=1.4\cdot 10^{-10}
\end{align}
for both sets of parameters. As discussed in the previous section, care has to be taken to use the right input parameters for each flavor.
\subsection{General discussion}

Averaging out the results from the two sets of parameters and using the spread as an estimate of the uncertainty, our final number for the contribution of Goldstone modes and axial-vector states is
\begin{align}\label{final}
a_{\mu}^{\rm{(AV+PS)}}=12.5(1.5)\cdot 10^{-10}\,.
\end{align}
The breakdown of the different contributions to this number is collected in Table~\ref{tab:1}. Our analysis shows that, since the axial-vector contribution is mostly constrained by the anomaly, the associated uncertainty happens to be small. The bulk of the uncertainty of our number comes instead from the Goldstone contribution, which can only be improved with better experimental data. However, this does not mean that our uncertainty on the axial-vector contribution is realistic. As already shown in \cite{Melnikov:2003xd}, the splitting of axial-vector masses inside the multiplet (something that our model cannot reproduce) can have a sizeable effect. This effect cannot be estimated within our model, but the observation recommends that we increase the axial-vector uncertainty.

Our result can be compared with previous literature on the same contributions, e.g.,
\begin{align}
a_{\mu}^{\rm{(AV+PS)}}&=13.6(1.5)\cdot 10^{-10}\,,\qquad &{\text{\cite{Melnikov:2003xd}}}\\
a_{\mu}^{\rm{(AV+PS)}}&=12.9(2.7)\cdot 10^{-10}\,,\qquad &{\text{\cite{Prades:2009tw}}}\\
a_{\mu}^{\rm{(AV+PS)}}&=12.1(2.1)\cdot 10^{-10}\,,\qquad &{\text{\cite{Jegerlehner:2009ry}}}\\
a_{\mu}^{\rm{(AV+PS)}}&=11.0(0.6)\cdot 10^{-10}\,,\qquad &{\text{\cite{Leutgeb:2019gbz}}}
\end{align} 
showing agreement. 

One might wonder whether the comparison above with~\cite{Melnikov:2003xd,Jegerlehner:2009ry,Prades:2009tw} makes sense. After all, the numbers found there were not obtained by resumming an infinite number of states, as we did in this work. However, those approaches obey the same short-distance constraints, which means that the numbers above are {\emph{effectively}} accounting for the same effects.

A different issue is how the results have to be interpreted. In particular, care has to be exercised when analyzing the relative weight of pseudoscalars and axial vectors in the numbers above. As we already discussed, our model shows that the MV limit is saturated by the longitudinal part of the axial-vector states. In the references quoted above, it was instead assumed (implicitly or explicitly) that pseudoscalars were saturating the constraint. 

In view of these differences, a more meaningful exercise would be to compare the longitudinal and transverse contributions of each determination separately. One would then find
\begin{align}
a_{\mu}^{L}&=11.1(1.5)\cdot 10^{-10}\,,
\end{align}
to be compared with 
\begin{align}
a_{\mu}^{\rm{(PS)}}&=11.4(1.0)\cdot 10^{-10}\,,\qquad &{\text{\cite{Melnikov:2003xd}}}\nonumber\\
a_{\mu}^{\rm{(PS)}}&=11.4(1.3)\cdot 10^{-10}\,,\qquad &{\text{\cite{Prades:2009tw}}}\nonumber\\
a_{\mu}^{\rm{(PS)}}&=9.9(1.6)\cdot 10^{-10}\,,\qquad &{\text{\cite{Jegerlehner:2009ry}}}
\end{align}
and
\begin{align}
a_{\mu}^{T}&=1.4(0.2)\cdot 10^{-10}\,,
\end{align}
to be compared with 
\begin{align}
a_{\mu}^{\rm{(AV)}}&=2.2(0.5)\cdot 10^{-10}\,,\qquad &{\text{\cite{Melnikov:2003xd}}}\nonumber\\
a_{\mu}^{\rm{(AV)}}&=1.5(1.0)\cdot 10^{-10}\,,\qquad &{\text{\cite{Prades:2009tw}}}\nonumber\\
a_{\mu}^{\rm{(AV)}}&=2.2(0.5)\cdot 10^{-10}\,.\qquad &{\text{\cite{Jegerlehner:2009ry}}}
\end{align}
The results show that the overall agreement carries over to the separate longitudinal and transverse contributions. However, when it comes to the separate particle contributions, we find
\begin{align}
a_{\mu}^{\rm{(PS)}}=9.6(1.6)\cdot 10^{-10}\,;\qquad  a_{\mu}^{\rm{(AV)}}=2.8(0.2)\cdot 10^{-10}\,.
\end{align} 
In other words, we find that the relative weight of axial vectors is substantially bigger than previously claimed in the literature. The same conclusion was also reached in~\cite{Leutgeb:2019gbz}. This does not mean that previous analyses should increase their estimate for axial vectors, as long as the MV constraint is satisfied. However, it shows that the analyses of the lowest-lying axial-vector contributions in~\cite{Jegerlehner:2017gek,Pauk:2014rta,Danilkin:2019mhd,Roig:2019reh} grossly underestimate the role of axial-vector contributions.

A more detailed analysis of the longitudinal piece shows that, despite the numerical agreement with~\cite{Melnikov:2003xd}, the relative increase in $a_{\mu}^{L}$ associated to states other than the Goldstone modes is more modest in our case, roughly $14\%$. Given that we satisfy the same short-distance constraints as~\cite{Melnikov:2003xd}, this difference has to have the origin in a different kinematic regime. In~\cite{Colangelo:2019lpu,Colangelo:2019uex} the impact of the MV constraint on the HLbL was estimated using a model with an infinite tower of pseudoscalars. A smaller number than the one in~\cite{Melnikov:2003xd} was also found, with the discrepancy identified by differences at low energies. Based on our conclusion in Sec.~\ref{sec:6} that axial-vector mesons saturate the MV constraint, we believe that massive pseudoscalars do not play a role in fulfilling the MV constraint (see also related comments in~\cite{Prades:2009tw}). Actually, the conceptual difficulties acknowledged in~\cite{Colangelo:2019uex} when taking the chiral limit are absent if axial vectors are considered. It is nevertheless instructive to compare their numerical results with ours. The comparison has to be made between our longitudinal axial-vector contribution and what~\cite{Colangelo:2019lpu,Colangelo:2019uex} define as their short-distance contribution. The reason is that the model used in~\cite{Colangelo:2019lpu,Colangelo:2019uex} can be taken simply as an interpolator between low and high energies, such that it effectively captures the same effects we have studied. With these caveats, we find good numerical agreement with them.   
           
\section{Conclusions}\label{sec:8}

The evaluation of the hadronic light-by-light contribution to the muon $(g-2)$ contains a number of conceptual issues which are hard to address using the approaches employed to date. For instance, a consistent phenomenological implementation of the so-called Melnikov-Vainshtein limit has turned out to be particularly challenging. The underlying problem is how to match OPE constraints with resonance exchanges, which happens to be highly nontrivial for the muon HLbL.

A framework suitable to study these issues should be able to evaluate hadronic effects from a Lagrangian formalism while being able to reproduce the right high-energy limits of QCD correlators. In other words, one would need a consistent realization of hadronic physics at the level of correlators.   

This can be done if one starts from a Lagrangian formulation in five dimensions and integrates out the fifth dimension. The spectrum of the resulting effective four-dimensional action contains an infinite number of resonances, with the quantum numbers of the fields introduced in the initial Lagrangian. In this paper we have chosen a minimal version of such constructions. The resulting four-dimensional effective action has a number of interesting features: (a) it is a theory of Goldstone modes consistently coupled to full towers of vector and axial-vector resonances; (b) the anomaly is consistently implemented at the hadronic level, i.e., at all energies; (c) the high-energy limit of correlators matches the pQCD predictions, such that quark-hadron duality is correctly implemented; and (d) it generates a phenomenologically successful pion transition form factor. 

With this toy model we have evaluated the contributions of pseudoscalar and axial-vector resonances to the HLbL four-point electromagnetic correlator in an inclusive way. We have thereby clarified why the phenomenological implementation of the Melnikov-Vainshtein limit at the hadronic level was elusive: the limit results from a collective effect of axial-vector resonances and, accordingly, cannot be reproduced with a finite number of states. Similar conclusions were recently drawn in~\cite{Leutgeb:2019gbz}. We have also explicitly shown how the sum rule that enforces the Melnikov-Vainshtein limit is the same that implements the anomaly in the $VVA$ triangle through a nontrivial interplay (a sum rule driven by anomaly matching) between pseudoscalars and the whole tower of axial vectors. 

Our final number for the joint Goldstone and axial-vector contributions is
\begin{align}
a_{\mu}^{\rm{HLbL, (PS+AV)}}=12.5(1.5)\cdot 10^{-10}\,,
\end{align}
where the uncertainty is simply orientative. 

This number agrees with previous estimates of the same contributions that took the MV constraint into account, but there are some important points to note. First, we claim a much bigger role for the axial-vector contribution, part of whose numerical impact has been commonly ascribed to excited pseudoscalar contributions in previous analyses. Second, for the relative weight that can be associated with the MV constraint compared to the Goldstone contribution, we find roughly $14\%$, which corresponds entirely to the longitudinal axial-vector contribution. This relative contribution is smaller than the one claimed in~\cite{Melnikov:2003xd} and comparable to the recent estimates in~\cite{Leutgeb:2019gbz,Colangelo:2019uex}. 

\section*{Aknowledgments}
We thank Marc Knecht for invaluable help and Massimilano Procura for useful discussions during the different stages of this work. We also thank Gilberto Colangelo, Franziska Hagelstein, Martin Hoferichter, Laetitia Laub, Josef Leutgeb, Anton Rebhan and Peter Stoffer for correspondence after the first version of this paper appeared. O.C. thanks the University of Naples and INFN-Sezione di Napoli for pleasant stays during the completion of this work. L.C., G.D. and A.I. were supported in part by MIUR under Project No. 2015P5SBHT and by the INFN research initiative ENP. The work of O.C. is supported in part by the Bundesministerium for Bildung und Forschung (BMBF FSP-105), and by the Deutsche Forschungsgemeinschaft (DFG FOR 1873). This research was supported by the Munich Institute for Astro- and Particle Physics (MIAPP), funded by the Deutsche Forschungsgemeinschaft (DFG, German Research Foundation) under Germany's Excellence Strategy --EXC-2094--390783311. 

\appendix

\section{Bulk-to-boundary propagators and Green functions}\label{app:1}

In order to apply the AdS/CFT prescription, it is convenient to split the fields in terms of the four-dimensional sources. For our purposes, this can be written as 
\begin{align}\label{fieldsAA}
A_{\mu}(q,z)&\equiv a(q,z){\hat{a}}_{\mu}^{\perp}(q)+{\bar{a}}(q,z){\hat{a}}_{\mu}^{\parallel}(q)+\alpha(z)iq_{\mu}\frac{\pi}{f_{\pi}}\,,\nonumber\\
V_{\mu}(q,z)&\equiv v(q,z){\hat{v}}_{\mu}^{\perp}(q)\,,
\end{align}
where $q$ is a four-dimensional momentum. The leading-order solutions for the functions $v(z,q)$, $a(z,q)$, ${\bar{a}}(z,q)$ are called the bulk-to-boundary propagators, which are determined from the solutions of the linearized equations of motion in five dimensions:
\begin{align}\label{difeq}
\left[\partial_z\left(\frac{1}{z}\partial_z\right)\eta_{\mu\nu}+\frac{q^2}{z}P^{\perp}_{\mu\nu}\right]v(z,q)=0\,,\nonumber\\
\left[\partial_z\left(\frac{1}{z}\partial_z\right)\eta_{\mu\nu}+\frac{q^2}{z}P^{\perp}_{\mu\nu}\right]a(z,q)=0\,,\nonumber\\
\left[\partial_z\left(\frac{1}{z}\partial_z\right)\eta_{\mu\nu}\right]{\bar{a}}(z,q)=0\,,
\end{align}
stemming from the quadratic part of the Yang-Mills term.  

The solutions for the transverse components can be determined once the boundary conditions are specified. Because of the previous factorization, at $z=0$ the mode functions get normalized to unity, $v(0,q)=a(0,q)={\bar{a}}(z,q)=1$.
At $z=z_0$ chiral symmetry breaking is enforced with (see the discussion in the main text) 
\begin{align}\label{VAbc}
\partial_zv(z_0,q)=0\,, \quad \; a(z_0,q)=0={\bar{a}}(z_0,q)\,.
\end{align}
For the evaluation of the different quantities we will mostly work in Euclidean space. The solutions in that case can be written in terms of modified Bessel functions as
\begin{align}
v(z,Q)&=Qz\big[K_1(Qz)+\xi_0 I_1(Qz)\big]\,,\label{vAdSE}\\
a(z,Q)&=Qz\big[K_1(Qz)-\xi_1 I_1(Qz)\big]\,,\label{aAdSE}
\end{align}
with 
\begin{align}
\xi_0=\frac{K_0(Q z_0)}{I_0(Q z_0)}\,;\qquad \xi_1=\frac{K_1(Q z_0)}{I_1(Q z_0)}\,.
\end{align}
The solution for the axial-vector longitudinal component can be found by taking the $Q\to 0$ limit of eq.~(\ref{aAdSE}). The result is 
\begin{align}
{\bar{a}}(z,q)=a(z,0)=1-\frac{z^2}{z_0^2}\equiv \alpha(z)\,.
\end{align}

The following identity:
\begin{align}\label{Wronskian}
v'(z,Q)a(z,Q)-v(z,Q)a'(z,Q)=zQ^2(\xi_0+\xi_1)\,,\nonumber\\
\end{align}
follows from the properties of the Wronskian of (\ref{difeq}). It is convenient to rewrite it in the form
\begin{align}
v'(z,Q)a(z,Q)-v(z,Q)a'(z,Q)=-\frac{z_0^2Q^2}{2}\alpha'(z)(\xi_0+\xi_1)\,,
\end{align} 
which is used in the main text to simplify a number of expressions.

The Green function for the axial-vector channel can be determined from the equations 
\begin{align}
\left[\partial_z\left(\frac{1}{z}\partial_z\right)+\frac{q^2}{z}\right]G^A_{\perp}(z,z^{\prime};q)&=\delta(z-z^{\prime})\,,\nonumber\\
\left[\partial_z\left(\frac{1}{z}\partial_z\right)\right]G^A_{\parallel}(z,z^{\prime};q)&=\delta(z-z^{\prime})\,,
\end{align}
where we have decomposed it in terms of its longitudinal and transverse projectors, i.e., 
\begin{align}
G_{\mu\nu}^A(z,z^{\prime};q)&= P^{\perp}_{\mu\nu}G_{\perp}^A(z,z^{\prime};q)+P^{\parallel}_{\mu\nu}G_{\parallel}^A(z,z^{\prime};q)\,.
\end{align} 
It satisfies the boundary conditions
\beq\label{bcGreen}
G_{\mu\nu}^A(0,z' ;q)=0\,, \quad G_{\mu\nu}^A(z_0,z';q)=0\,,
\eeq
and the following continuity conditions at $z=z'$:
\begin{align}\label{matchingGreen}
&G_{\mu\nu}^A(z'+\epsilon,z' ;q)=G_{\mu\nu}^A(z'-\epsilon,z' ;q)\,,\nonumber\\
&\partial_zG_{\mu\nu}^A(z'+\epsilon,z' ;q)- \partial_zG_{\mu\nu}^A(z'-\epsilon,z' ;q)=z'\,.
\end{align}
Since 
\begin{align}
G_{\parallel}^j(z,z')=G_\perp^j(z,z';0)\,,
\end{align}
it suffices to work out the solution of the transverse component. In Euclidean space, it is given by
\begin{equation}\label{GFA}
G^A_\perp(z,z' ;Q)=\left\{
\begin{array}{cc}
-\displaystyle\frac{(v(z',Q)-a(z',Q))\,a(z,Q)}{Q^2(\xi_0+\xi_1)},& z>z'\\
	\\
- \displaystyle\frac{a(z',Q)\,(v(z,Q)-a(z,Q))}{Q^2(\xi_0+\xi_1)},& z<z'\,.
\end{array}
\right.
\end{equation}
Using the results of (\ref{vAdSE}) and (\ref{aAdSE}), the expression can be simplified to 
\begin{equation}\label{GAT}
G^A_{\perp}(z,z^{\prime};Q)=-\frac{1}{Q^2}\left\{
\begin{array}{cc}
\displaystyle Qz^{\prime}I_1(Qz^{\prime})a(z,Q),& z>z^{\prime}\\
\\
\displaystyle a(z^{\prime},Q)QzI_1(Qz),& z<z^{\prime}\,.
\end{array}
\right.
\end{equation}
The longitudinal component is obtained as 
\begin{equation}
G_A^L(z,z^{\prime};Q)=G_A^{\perp}(z,z^{\prime};0)=\left\{
\begin{array}{cc}
-\displaystyle \frac{(z^{\prime})^2}{2}\alpha(z)& z>z^{\prime}\\
\\
-\displaystyle \frac{z^2}{2}\alpha(z^{\prime})& z<z^{\prime}\,.
\end{array}
\right.
\end{equation}
Using the expression (\ref{GFA}) and the relation (\ref{Wronskian}), one obtains that
\begin{widetext}
\begin{align}
\partial_z\partial_{z'}G_A^{\perp}(z,z',Q)&=\frac{z_0^2}{2}\alpha'(z)\delta(z-z')+\frac{1}{Q^2(\xi_0+\xi_1)}\bigg[a'(z)a'(z')-v'(z')a'(z)\theta(z-z')-v'(z)a'(z')\theta(z'-z)\bigg]\,.
\end{align}
\end{widetext}
The longitudinal component is the zero-momentum limit of the above expression, which leads to
\begin{align}
\partial_z\partial_{z'}G_A^{\parallel}(z,z')=\frac{z_0^2}{2}\bigg[\alpha'(z)\delta(z-z')+\alpha'(z)\alpha'(z')\bigg]\,.
\end{align}

\section{Axial-vector two-point correlator}\label{app:2}

The computation of the two-point function
\begin{align}
\Pi^{AA}_{\mu\nu}(q)&=i\int d^4x e^{iq\dot x}\langle 0|T\left\{J_{\mu}(x)J_{\nu}(0)\right\}|0\rangle
\end{align}
is a simple example to illustrate the importance of keeping track of sources in order to have consistent expressions for correlators. It will also be used to fix the values of the free parameters of the model. Our results will be in the exact chiral limit. 

The relevant part of the action to compute the correlator is the quadratic term in axial-vector sources. In the holographic prescription, this term is given by
\begin{align}\label{SAA}
S_{\mathrm{eff}}&=-2\lambda \int d^4x A_{\mu}(x)\frac{1}{z}\partial_z A^{\mu}(x)\bigg|_{z=0}\,.
\end{align} 
Plugging eq.~(\ref{fieldsAA}) above, one finds contact terms for the perpendicular and longitudinal axial-vector sources, together with a vertex connecting the pion to the longitudinal axial-vector source. Notice this important feature of the model: while the pion and axial-vector modes are decoupled (this determines the form of $\alpha(z)$), the pion still couples to the longitudinal axial-vector {\it{sources}}. This just shows that the model correctly implements PCAC. 

The explicit expression for the two-point correlator takes the form
\begin{widetext}
\begin{align}\label{HAA}
\Pi^{AA}_{\mu\nu}(q)&=-4\lambda\left[P^{\perp}_{\alpha\mu}P^{\perp}_{\alpha\nu} a(q,z)\frac{1}{z}\partial_z a(q,z)+P^{\parallel}_{\alpha\mu}P^{\parallel}_{\alpha\nu} {\bar{a}}(q,z)\frac{1}{z}\partial_z {\bar{a}}(q,z)+P^{\parallel}_{\alpha\mu}P^{\parallel}_{\alpha'\nu} \frac{q^{\alpha}q^{\alpha'}}{q^2}\alpha(z)\frac{1}{z}\partial_z \alpha(z)\right]\bigg|_{z=0}\nonumber\\
&=-4\lambda\left[P^{\perp}_{\mu\nu}a(q,z)\frac{1}{z}\partial_z a(q,z)+P^{\parallel}_{\mu\nu}{\bar{a}}(q,z)\frac{1}{z}\partial_z {\bar{a}}(q,z)+P^{\parallel}_{\mu\nu}\alpha(z)\frac{1}{z}\partial_z \alpha(z)\right]\bigg|_{z=0}\,,
\end{align}
\end{widetext}
where the last term corresponds to the pion propagation.

At very low energies one can check that $a(0,z)={\bar{a}}(0,z)=\alpha(z)$ and the expression above can be simplified to
\begin{align}\label{chpt}
\lim_{q^2\to 0}\Pi^{AA}_{\mu\nu}(q)&=-4\lambda (P^{\perp}_{\mu\nu}+2P^{\parallel}_{\mu\nu})\alpha(z)\frac{1}{z}\partial_z \alpha(z)\bigg|_{z=0}\nonumber\\
&=\frac{8\lambda}{z_0^2} P^{\perp}_{\mu\nu}\,,
\end{align} 
where in the last line we have used the explicit expression for $\alpha(z)$. The previous equation can be alternatively obtained using that
\begin{align}\label{chiralA}
A_{\mu}(0,x)=\frac{u_{\mu}}{2}\equiv -\frac{i}{2}u^{\dagger}D_{\mu}U u^{\dagger}\,,
\end{align}
where $D_{\mu}U=\partial_{\mu}U-il_{\mu}U+iUr_{\mu}$. This ensures, in particular, that the low-energy limit of this model matches chiral perturbation theory, as it should. This means that the prefactor in (\ref{chpt}) is to be identified with $f_{\pi}^2$, i.e.,
\begin{align}
\frac{8\lambda}{z_0^2}=f_{\pi}^2\,,
\end{align}
and thus one obtains eq.~(\ref{fpi}) in the main text.  

The important point to emphasize is that the pion propagation alone is longitudinal, and only the inclusion of {\it{both}} the longitudinal and the transversal local contact terms makes $\Pi^{AA}_{\mu\nu}$ a transverse object, as required by the Ward identity in the chiral limit. This role of the contact terms is in close analogy to what we observed for the $VVA$ correlator in Sec.~\ref{sec:5}.

The leading short-distance behavior of eq.(\ref{HAA}) can be found by expanding the two-point axial-vector correlator close to the UV boundary. 
\begin{align}
\lim_{Q^2\to \infty}\Pi^{AA}_{\mu\nu}(q)&=-2\lambda \log\frac{Q^2}{\mu^2} P^{\perp}_{\mu\nu}\,,
\end{align}
which is reproduced from the transverse piece alone. Matching to the coefficient of the parton-model logarithm gives the result reported in eq.~(\ref{g5Nc}).  
  
\section{Scalar coefficients from the transverse part of the axial-vector Green function}\label{app:3}

If one uses the formalism described in \cite{Colangelo:2015ama}, one can decompose the HLbL tensor into 54 tensorial structures. Details and definitions can be found in this reference. In this Appendix we just list the form of the scalar functions ${\bar\Pi}_i(q_1^2,q_2^2, q_3^2),\; i=1,...12$ as predicted by our model. 

Defining 
\begin{align}
{\mathcal W}^T&(q_1^a,q_b^2; q_c^2,q_d^2)\equiv\int_0^{z_0}dz\int_0^{z_0}dz' v(z,q_a^2)\partial_z v(z,q_b^2)\nonumber\\
&\nonumber\\
&\times G_A^T(z,z';(q_a+q_b)^2)v(z',q_c^2)\partial_{z'}v(z',q_d^2)\,,
\end{align}
one  gets ($q_4=0$ and $q_3=-q_1-q_2$)
\begin{widetext}
\begin{align}\label{12Apostoles}
{\bar\Pi}_3(q_1^2,q_2^2,q_3^2)&=\frac{1}{q_1^2 q_2^2(q_1+q_2)^2}\left[q_2^2(q_1\cdot q_2+ q_2^2) {\mathcal W}^T\left(q_2^2,q_3^2;0,q_1^2\right)+q_1^2\left(q_1\cdot q_2+ q_1^2\right){\mathcal W}^T\left(0,q_2^2;q_1^2,q_3^2\right)\right]\,, 
\\
{\bar\Pi}_4(q_1^2,q_2^2,q_3^2)&=\frac{1}{q_1^2 q_2^2(q_1+q_2)^2}\left[q_3^2(q_1\cdot q_2+ q_2^2) {\mathcal W}^T(q_3^2,q_2^2;0,q_1^2)-q_1^2\;q_1\cdot q_2{\mathcal W}^T(q_1^2,q_2^2;0,q_3^2)\right]\,,
\\
{\bar\Pi}_5\left(q_1^2,q_2^2,q_3^2\right)&=\frac{1}{{q_1^2} {q_2^2} \left(q_1+q_2\right)^2}\left[{q_2^2} {\mathcal W}^T\left(q_2^2,q_3^2;0,q_1^2\right)-\left(q_1+q_2\right)^2 {\mathcal W}^T\left(q_3^2,q_2^2;0,q_1^2\right)-{q_1^2} {\mathcal W}^T\left(0,q_2^2;q_1^2,q_3^2\right)\right]\,, 
\\
{\bar\Pi}_6\left(q_1^2,q_2^2,q_3^2\right)&=\frac{1}{{q_1^2} {q_2^2} \left(q_1+q_2\right)^2}\left[{q_1^2} {\mathcal W}^T\left(q_1^2,q_2^2;0,q_3^2\right)-{q_2^2} {\mathcal W}^T\left(q_2^2,q_1^2;0,q_3^2\right)-\left(q_1+q_2\right)^2 {\mathcal W}^T\left(q_3^2,q_2^2;0,q_1^2\right)\right]\,,
\\
{\bar\Pi}_7\left(q_1^2,q_2^2,q_3^2\right)&=-\frac{1}{{q_1^2} {q_2^2} \left(q_1+q_2\right)^2}\left[{q_1^2} {\mathcal W}^T\left(q_1^2,q_2^2;0,q_3^2\right)+{q_2^2} {\mathcal W}^T\left(q_2^2,q_3^2;0,q_1^2\right)-\left(q_1+q_2\right)^2 {\mathcal W}^T\left(q_3^2,q_2^2;0,q_1^2\right)\right]\,,
\\
{\bar\Pi}_8\left(q_1^2,q_2^2,q_3^2\right)&= \frac{1}{2{q_1^2} {q_2^2} \left(q_1+q_2\right)^2}\left[{q_2^2}\left( {\mathcal W}^T\left(q_2^2,q_3^2;0,q_1^2\right)-{\mathcal W}^T\left(q_2^2,q_1^2;0,q_3^2\right)\right)\right.\nonumber\\
&\!\!\!\!\!\!\!\!\!\!\!\!\left.+{q_1^2} \left({\mathcal W}^T\left(q_1^2,q_2^2;0,q_3^2\right)+{\mathcal W}^T\left(0,q_2^2;q_1^2,q_3^2\right)\right)+\left(q_1+q_2\right)^2 \left({\mathcal W}^T\left(q_3^2,q_2^2;0,q_1^2\right)-{\mathcal W}^T\left(0,q_2^2;q_3^2,q_1^2\right)\right)\right]\,,
\\
{\bar\Pi}_9\left(q_1^2,q_2^2,q_3^2\right)&= \frac{1}{2{q_1^2} {q_2^2} \left(q_1+q_2\right)^2}\left[{q_1^2}\left( {\mathcal W}^T\left(q_1^2,q_2^2;0,q_3^2\right)- {\mathcal W}^T\left(0,q_2^2;q_1^3,q_3^2\right)\right)\right.\nonumber\\
&\!\!\!\!\!\!\!\!\!\!\!\!\left.+{q_2^2}\left( {\mathcal W}^T\left(q_2^2,q_1^2;0,q_3^2\right)- {\mathcal W}^T\left(q_2^2,q_3^2;0,q_1^2\right)\right)+\left(q_1+q_2\right)^2 \left({\mathcal W}^T\left(q_3^2,q_2^2;0,q_1^2\right)+{\mathcal W}^T\left(0,q_2^2;q_3^2,q_1^2\right)\right)\right]\,,
\\
{\bar\Pi}_{10}\left(q_1^2,q_2^2,q_3^2\right)&= \frac{1}{2{q_1^2} {q_2^2} \left(q_1+q_2\right)^2}\left[{q_2^2} \left({\mathcal W}^T\left(q_2^2,q_1^2;0,q_3^2\right)+{\mathcal W}^T\left(q_2^2,q_3^2;0,q_1^2\right)\right)\right.\nonumber\\
&\!\!\!\!\!\!\!\!\!\!\!\!\left.+{q_1^2} \left({\mathcal W}^T\left(q_1^2,q_2^2;0,q_3^2\right)+{\mathcal W}^T\left(0,q_2^2;q_1^2,q_3^2\right)\right)+\left(q_1+q_2\right)^2\left({\mathcal W}^T\left(q_3^2,q_2^2;0,q_1^2\right)+{\mathcal W}^T\left(0,q_2^2;q_3^2,q_1^2\right)\right)\right]\,,
\\
{\bar\Pi}_{11}\left(q_1^2,q_2^2,q_3^2\right)&= \frac{1}{2{q_1^2} {q_2^2} \left(q_1+q_2\right)^2} \left[{q_1^2}\left( {\mathcal W}^T\left(q_1^2,q_2^2;0,q_3^2\right)- {\mathcal W}^T\left(0,q_2^2;q_1^2,q_3^2\right)\right)\right.\nonumber\\
&\!\!\!\!\!\!\!\!\!\!\!\!\left.+{q_2^2}\left( {\mathcal W}^T\left(q_2^2,q_1^2;0,q_3^2\right)-{\mathcal W}^T\left(q_2^2,q_3^2;0,q_1^2\right)\right)-\left(q_1+q_2\right)^2\left( {\mathcal W}^T\left(q_3^2,q_2^2;0,q_1^2\right)- {\mathcal W}^T\left(0,q_2^2;q_3^2,q_1^2\right)\right) \right]\,,
\\
{\bar\Pi}_{12}\left(q_1^2,q_2^2,q_3^2\right)&= \frac{1}{2{q_1^2} {q_2^2} \left(q_1+q_2\right)^2}\left[{q_1^2} \left({\mathcal W}^T\left(q_1^2,q_2^2;0,q_3^2\right)-{\mathcal W}^T\left(0,q_2^2;q_1^2,q_3^2\right)\right)\right.\nonumber\\
&\!\!\!\!\!\!\!\!\!\!\!\!\left.-{q_2^2} \left({\mathcal W}^T\left(q_2^2,q_1^2;0,q_3^2\right)- {\mathcal W}^T\left(q_2^2,q_3^2;0,q_1^2\right)\right)+\left(q_1+q_2\right)^2\left( {\mathcal W}^T\left(q_3^2,q_2^2;0,q_1^2\right)- {\mathcal W}^T\left(0,q_2^2;q_3^2,q_1^2\right)\right)\right]\,.
\end{align}
\end{widetext}
In all the  $\bar\Pi_i$ above,  $v(z,q_4^2)$ (or $v(z',q_4^2)$ ) always appear without derivative, and since we are taking the $q_4\rightarrow 0$ limit, if follows that $v(z,q_4^2)\rightarrow 1$. Thus, the integral at the soft-photon vertex contains the product of the transverse axial-vector Green function and only one derivative both depending on the same four-momentum. This leads to simplifications.

For instance, let us consider the case
\begin{align}
{\mathcal W}^T(q_1^2,q_2^2; 0,q_3^2)&=\int_0^{z_0}\;dz\int_0^{z_0}\;dz' v(z,q_1^2)\partial_zv(z,q_2^2)\nonumber\\
&\times G_A^T(z,z';q_3^2)\partial_{z'}v(z',q_3^2)\,,
\end{align}
where $q_3=-q_1-q_2$, and we have used that $v(z,0)=1$. Then the following identity holds:
\begin{align}
&\int_0^{z_0} dz'\,G_A^T(z,z';q^2)\partial_{z'}v(z',q^2)=\nonumber\\
&\qquad \frac{1}{2}\frac{z_0^2}{2}\left(a(z,q^2)-\alpha(z)v(z,q^2)\right)\,,
\end{align}\label{Wronski}

so that one is left with a single integration:
\begin{align}
&{\mathcal W}^T(q_1^2,q_2^2; 0,(q_1+q_2)^2)=\frac{1}{2}\frac{z_0^2}{2}\int_0^{z_0}\;dz\, v(z,q_1^2)\nonumber\\
&\qquad\times\partial_zv(z,q_2^2) \left(a(z,(q_1+q_2)^2)-\alpha(z)v(z,(q_1+q_2)^2)\right)\,,\label{SingleIntGT}
\end{align}
which can be evaluated numerically.

\end{document}